%Paper: hep-th/9403075
%From: "V.B. Petkova" <PTVP@IBM.RZ.TU-CLAUSTHAL.DE>
%Date: Mon, 14 Mar 94 20:44:14 MET
%Date (revised): Sun, 20 Mar 94 18:34:15 MET
%Date (revised): Sun, 25 Sep 94 19:26:38 MET

\magnification 1100

%%%%%%%%%%%%%%
%
%   macros
%
%%%%%%%%%%%%%%

\def\Tf{T^{\rm f\,f}}

\def\za{\alpha} \def\zb{\beta}  \def\zd{\delta}
\def\ze{\varepsilon}   \def\zk{\kappa}
\def\zl{\lambda} \def\zm{\mu} \def\zn{\nu} 
 \def\zr{\rho}

\def\zO{\Omega}

% caligraphic
%

  \def\cL{{\cal L}} 
  \def\cO{{\cal O}}  \def\cP{{\cal P}}
    \def\cS{{\cal S}}
   \def\cV{{\cal V}}
\def\cW{{\cal W}}

\def\Ri{{\cal R}^{(i)}}
\def\odots{{\scriptstyle{\circ\atop\circ}}}
 
\def\p{\partial}

 \def\H{h} \def\C{C}
\def\kk{{1\over\nu}}
\def\Tf{T^{(\rm f{}f)}} \def\Wf{W^{(\rm f{}f)}}
\def\Lf{L^{(\rm f{}f)}}
\def\aw{a_{\nu}}

\def\HH{H}
\def\hh{\hat h} \def\hf{\hat f} \def\he{\hat e}
\def\dZ{Z\!\!\!Z}\def\R{{\cal R}}

\def\hs{\widehat {sl}(3)}
\def\V{V_{\lambda}}
\def\Vi{V^{(i)}}
\def\W{{\cal W}}

\def\Vm{V_\mu}
\def\fb{{\bf f}}

\def\f10{ f^1_0}
\def\e10{ e^1_0}
\def\Xf{{\cal X}^{\rm (f{}f)}}
\def\cn{{C_{11}\over\nu}}

% doubled letters:

   \def\dC{I\!\!\!\!C}

 \def\dN{I\!\!N}
       
     \def\dZ{Z\!\!\!Z}

\def\la{\langle} \def\ra{\rangle}

\def\db{{\bf d}} \def\eb{{\bf e}} \def\fb{{\bf f}}
\def\kb{{\bf k}} \def\hb{{\bf h}}
\def\pb{{\rm (PB)}} 
\def\ol{\overline\lambda}

%%%%%%%%%%%%%%%%%
%%                                     %%
%%    title page                   %%
%%                                      %%
%%%%%%%%%%%%%%%%%%

\nopagenumbers
\hfuzz=20pt

\centerline{ }
\hfill                                         INFN/AE-94/10

\hfill                                         KL-TH-94/6

\hfill                                         ASI-TPA/6/94

\hfill                                     {\tt hep-th/9403075}

\vskip 2cm
\centerline{\bf  Singular Vectors of ${\cal W}$ Algebras
            via DS Reduction of ~$A_2^{(1)}$}

\vskip 2cm

\centerline{
{\bf P. Furlan${}^{\dagger +}$},~
{\bf A.Ch. Ganchev${}^{\dagger\sharp *\flat}$}~ and ~
{\bf V.B. Petkova${}^{\dagger \star *}$}}
\vskip 1cm

\hskip 1cm
\vbox{
\item{${}^\dagger$} Istituto Nazionale di Fisica Nucleare,
Sezione di Trieste,               Italy
                                  \medskip
\item{${}^+$}
Dipartimento di Fisica Teorica
              dell'Universit\`a di Trieste, Italy
        \medskip
\item{${}^\sharp$}          Fachbereich Physik, Kaiserslautern
University,
              Germany
        \medskip
\item{${}^\star$} Arnold Sommerfeld Institute for Mathematical
Physics,
\hfil\break
         Technical University Clausthal,
Germany }

\vskip 2cm

\centerline{\bf Abstract}
\vskip 1cm

The  BRST quantisation of the Drinfeld - Sokolov reduction
applied to the case of $A^{(1)}_2\,$ is explored to construct in
an unified and systematic way the general singular vectors in
${\cal W}_3$ and ${\cal W}_3^{(2)}$ Verma modules.  The
construction relies on the use of proper quantum analogues of the
classical DS gauge fixing transformations.  Furthermore the
stability groups  $\overline W^{(\eta)}\,$ of the highest weights
of the ${\cal W}\,$  -  Verma modules play an important role in
the proof of the BRST equivalence of the Malikov-Feigin-Fuks
singular vectors  and the  ${\cal W}$ algebra ones.  The
resulting singular vectors are essentially classified by the
affine Weyl group $W\, $  modulo  $\overline W^{(\eta)}\,$.

This is a detailed presentation of the results announced in a
recent paper of the authors.

\vskip 1cm

\footnote{}{
{${}^*$}\   Permanent address:  Institute for Nuclear
          Research and Nuclear Energy, Sofia, Bulgaria
\smallskip
{}\ {${}^\flat$}\       Humboldt fellow           }

\vfil March 1994

\vfill\eject
\pageno=1
\footline{\hfil\tenrm\folio\hfil}

%%%%%%%%%%%%%%
%
%   section 1.
%
%%%%%%%%%%%%%%%%

\bigskip\noindent{\bf 1. INTRODUCTION.} \medskip

The Hamiltonian reduction of affine Lie algebras (the
Drinfeld-Sokolov (DS) reduction) is a  powerful method for
obtaining and analysing ${\cal W}$ algebras.  Its classical
aspects are well understood [1], [2], [3], [4], [5]  -- starting
from a classical analogue of a Kac-Moody (KM) algebra $\hat g$
one imposes  on the raising operators constraints specified by a
$sl(2)$ embedding in $g$. The factorisation over the gauge group
generated by the (first class) constraints yields a set of
invariant differential polynomials of the affine generators.
These are the generating currents of the ${\cal W}$ (Poisson)
algebra.

The quantisation of the DS reduction is usually done in a BRST
framework and was originally carried  out using free field
realisations for the affine algebras and Fock modules [6], [7],
[8], [9].  The free field representation of KM and ${\cal W}$
algebras is quite involved in general and the representation
theory of the general ${\cal W}$ algebras is in a less developed
stage.

Recently de Boer and Tjin [10,11] proposed a different BRST
scheme for the reduction of an  affine Lie algebra
based, as in the classical case, on  $sl(2)$ embeddings and not
relying on the use of free fields. (A free field realisation of
the resulting ${\cal W}$ -algebra
appears nevertheless naturally.)
 This approach is suitable for the reduction of Verma module
representations.  The  description of the irreducible highest
weight representations  through Verma modules  has a
direct relevance for the underlying Conformal Field Theory
models. It provides  information on the characters
[12],  [13],
 and, accounting for the explicit form of the Verma module
singular vectors -- eventually on the differential and algebraic
equations for the  correlation functions.  Thus the further
development of the  approach in [11] would allow to carry over
the reduction procedure to the full conformal field theories, as
has been done in [14] for the simplest case $\widehat sl(2)
\rightarrow $ Virasoro algebra.

The description of the singular vectors of Kac-Moody
Verma modules is transparent and relatively simple [15], [16].
Thus it would be
desirable to have a method reducing these singular vectors to
their ${\cal W}$ algebra counterparts. This has been done in the
simplest case of $A_1^{(1)}$ in [17], recovering  the general
Virasoro Verma modules singular vectors in the form proposed by
Kent [18]. The idea underlying the derivation in [17] (see also
the earlier versions in [14], [19]) is to exploit a kind of a
quantum  analogue of the classical Drinfeld-Sokolov  gauge fixing
transformation, taken in an arbitrary representation.
In the classical case it amounts (constraining the
raising generator density $e(z)=1$)
to the mapping of the lowering generator $f(z)$
into the Virasoro stress-energy tensor,

$$\eqalign{
  & f(z) \rightarrow {1\over \nu}\,T(z)
    = f(z) + {1\over \nu}\, \Tf(z) \,,
\cr\cr
  & {1\over \nu}\,\Tf(z) = \Big( {h(z)\over 2}\Big)^2
 + \Big( {1\over \nu} - 1 \Big) \, {\partial\,h(z)\over 2} \,.
\cr}\eqno(1.1)$$
The Heisenberg subalgebra density $h(z)/2\,$ serves as a gauge
parameter.

The quantum BRST invariant  Virasoro stress-energy tensor
$T(z)=\sum z^{-n-2}\, L_n\, $ is given by a formula analogous to
that in (1.1), shifting $h(z)$ by a
ghost bilinear combination $\,\hat h(z)= h(z) + (bc)(z)\,,$
adding normal products and identifying the constant $1/\nu -2$
with the value $k$ of the $A_1^{(1)}$ level.  A
quantum analogue of the transformation (1.1) is provided by the
operator
$$\eqalign{
   {\cal R} &= {\bf 1} + {\hat h_{-1}\over 2}\, e_0
              + {1\over 2}  \Big( \big( {\hat h_{-1}\over 2}
\big)^2
                 - {\hat h_{-2}\over 2} \Big)\, e_0^2 + \dots
\cr\cr
         &= \sum_{k=0}^{\infty}
                 {1\over k!}\,{\partial^k\over\partial u^k}\,
              \left.\left(
              \exp \int_0^u {\hat h_{(-)} (-u')\over 2} \, du'
              \,\right)\right|_{u=0}\, e_0^k\,,\qquad
     \hat h_{(-)} (u)\equiv \sum_{n=1}^{\infty} u^{n-1}\,
	 \hat h_{-n}\,,
\cr}\eqno(1.2)$$
which has the following
properties:

\item{$\bullet\quad$}
{\it It leaves invariant the Kac-Moody singular vectors,}

\item{$\bullet\quad$}
{\it maps horizontal vectors into the kernel of the BRST  operator,}

\noindent
i.e.,  $Q$ kills states of the form ${\cal R}\,(f_0)^n\,V_\lambda$,
$n$ -  non-negative integer, $V_\lambda$ -  a Kac-Moody
highest weight state. Moreover and most important

\item{$\bullet\quad$}
{\it ${\cal R}$ intertwines Kac-Moody and Virasoro generators,}

\noindent
more precisely, for
any vector  $V$ annihilated by all positive grade generators as
well as by the ghost zero mode $c_0\,$, one has
$$
   {\cal R}\, f_0\, V = {1\over \nu } \sum_{p=0}
                            \,L_{-p-1} \,{\cal R} \,\, (-e_0)^p\,
V \,.
\eqno(1.3)$$
These properties together with the quantum constraint allow to
recover all the Virasoro singular vectors starting from proper
$\widehat{sl}(2)$ ones.

Generalising the above approach to the $\widehat{sl}(3)\,$ case,
we have outlined in the short letter [20] a program for the
construction of the general singular vectors in  Verma modules of
the related ${\cal W}\,$ algebras.  This work provides a detailed
presentation of the results announced in [20].

The paper is organised as follows. Section 2 contains notation
conventions and a summary of known facts about the reducibility
conditions of affine Kac-Moody Verma modules [15], [16].  More
details  on the Malikov-Feigin-Fuks (MFF) $A_2^{(1)}$ singular
vectors and an algorithm transforming Malikov-Feigin-Fuks
monomials into ordinary integer-powers polynomials are given in
Appendix A.  In section 3 we recall the basics about the
Zamolodchikov (Z) [21] algebra ${\cal W}_3$  and the Polyakov -
Bershadsky  (PB)  [22], [9] ${\cal W}_3^{(2)}$ algebras,
following the realisation of [11].  Section  4 is devoted to the
introduction of the quantum gauge operators relevant for the two
algebras  and the study of their properties.  Appendix B contains
the technical details of the derivation of the basic intertwining
relations. For comparison with the classical cases we refer the
reader to [20] and the original works [3], [4].  In Section 5 the
quantum gauge transformations are used to reduce the MFF
$A_2^{(1)}$ singular vectors to the singular vectors of ${\cal
W}_3^{(2)}$ Verma modules. Section 6 deals with the same problem
in the case of  ${\cal W}_3\,$, starting  with the subclass
of singular vectors first obtained by Bowcock and Watts [23],
following  the fusion method of [24].  Appendix C contains some
details of the argumentation leading to the general   ${\cal
W}_3$ singular vectors.  The symmetry groups of the eigenvalues
of the zero modes subalgebras play in both cases essential role,
determining the way the quantum constraints are accounted for in
the reduction  of the MFF vectors.
Since we start from the  MFF
vectors, the resulting ${\cal W}_3$ singular vectors are in a
form  analogous to the one proposed by Kent [18] for the Virasoro
case.  In Appendix D  we describe on an example how one can
obtain explicit expressions for the singular vectors starting
from Kent type of expressions.  We end with a discussion of open
problems.

%%%%%%%%%%%%%%%
%
%  section 2.
%
%%%%%%%%%%%%%%%

\bigskip\noindent{\bf 2. NOTATION AND SUMMARY OF KNOWN FACTS.}
\medskip
\noindent{\bf 2.1. Singular vectors of $A^{(1)}_2$ Verma modules}
\medskip

The positive roots of $\bar g = A_2$ consist of  the simple roots
$\alpha^1$, $\alpha^2$ and the highest root
 $\alpha^3 \equiv \alpha^1+\alpha^2$.
There is a nondegenerate form $\la\cdot,\cdot\ra$ on the space
spanned by the roots. The Cartan matrix
$\C^{ij}=\la\alpha^i,\alpha^j\ra$, $i,j=1,2$ and its
inverse $\C_{ij}$ in the case of $A_2$ are
$$
  (\C^{ij}) = \left( \matrix{ 2 &-1 \cr -1 &2 \cr} \right),
\qquad
  (\C_{ij}) = {1\over 3} \left(\matrix{ 2 &1 \cr 1 &2
\cr}\right)\,.
\eqno(2.1) $$
To each positive/negative root $\pm\alpha^i$
corresponds a raising/lowering operator
$e^i\, /\, f^i\,$, respectively.
To each simple root corresponds a Cartan generator $\H^i$,
$i=1,2\,,$
$\za^i (h^j)= C^{ij}\,.$
The commutation relations of $A_2$ are
$$\eqalign{
  [e^{i}, f^{j}] &= \delta_{ij} \, \H^i \,,
  \qquad [\H^i, \H^j] = 0 \,,   \qquad i=1,2\,, \cr\cr
  [\H^i, e^j] &=  \la\alpha^i,\alpha^j\ra\, e^j \,,
  \qquad [\H^i, f^j] = - \la\alpha^i,\alpha^j \ra\, f^j \,,
\cr}\eqno(2.2a) $$
and setting
$$      e^{3} = [e^{1},e^{2}]\,, \qquad
  f^{3} = [f^{2},f^{1}]\,,
\eqno(2.2b) $$
the other commutation relations are determined by the Jacobi
identity.
The quadratic form on the Cartan algebra can be extended to a
nondegenerate
symmetric form on the whole algebra by
$$
  \la \H^i,\H^j \ra = C^{ij}\,, \quad i,j=1,2
    \qquad  \la e^{a}, f^{a} \ra = 1\,,\quad a=1,2,3,
\eqno (2.3)$$
the others being zero.

Given the commutators $[X,Y]$ of the elements $X$, $Y$ of a
semisimple
Lie algebra $\bar g$ and the Cartan - Killing form
$\la X,Y \ra$ the affine algebra
$g$
 associated to $\bar g$
has commutation relations
$$
   [ X_n,\, Y_m ] = ( [ X,\, Y ] )_{n+m}
   + \kb \, n\, \delta_{n+m,0}\, \la X,Y\ra
\eqno(2.4) $$
where $\kb$ is a central element. The bilinear form is extended
to $g$ by $\la X_n,Y_m \ra=\zd_{n+m,0}\,\la X,Y\ra $ and
trivially to $\kb\,,$
i.e., $\la X_n,\kb \ra=0$.
The  algebra $g\,$ admits the decomposition $g= {\tt n_+}
\oplus {\tt h} \oplus {\tt n_-} \,,$ where ${\tt h}=\bar {{\tt
h}}\oplus
\dC\kb \,$ is the  Cartan subalgebra and the subalgebra  ${\tt
n_+}$ (${\tt
n_-}$) is generated,  via multiple commutators,
by
$$\eb_0=f^3_1\,, \eb_1=e^1_0\,,  \eb_2=e^2_0\,, \quad
  (\fb_0=e^3_{-1} \,,  \fb_1=f^1_0\,, \fb_2=f^2_0) $$
respectively.
These are the raising (lowering) operators
 corresponding to the simple roots  $\alpha^0 $,
$\alpha^1$,  $\alpha^2$. Denote
$\hb^0=[\eb_0, \fb_0 ]= \kb - h^3_0\,, \hb^i = h_0^i\,, \
i=1,2,$~ $h_0^3 = h_0^1 +h_0^2$. The Cartan matrix  of the
horizontal (zero mode) subalgebra  $\bar g$  extends to
$ \la \za^0, \za^j \ra
= \za^0( \hb^j) = \za^j(\hb^0) = 2\zd_{j0}-\zd_{j1}-\zd_{j2}\,.$

The set of real positive roots is
$$\Delta_{+,{\rm real}} =
\left\{ \alpha^i,\ \pm\alpha^i+n\zd
 : \ i=1,2,3;\ n\in\dN \right\}\,,
\eqno(2.5)$$
where $\zd \equiv \za^0 +  \za^3$.
The positive roots consist of the positive real roots and the
positive degenerate roots
 $\left\{ n\,
\zd :\ n\in\dN \right\}$ which have zero norm.

Denote by $\zl_0$ the element in $ {\tt h^*}\,,$ dual  to
$\kb\,,$
i.e., $\zl_0(\kb)=1 \,,$  $ \,  \zl_0( \hb^i)$ $ = 0\,, i=1,2$
and let  $ \la  \zl_0, \za^j \ra= \zl_0( \hb^j) \,, j=0,1,2,\,\,
\la \zl_0,\zl_0\ra =0.$
  It is convenient to enlarge the Cartan subalgebra
${\tt h}$
by adjoining a derivation $\db\,,$
 $[ \db, X_m ] = m\,X_m$ and $[ \db, \kb ]=0$.
 The bilinear  form on $g$ is extended to a nondegenerate one
by requiring $\la \db, \kb \ra=1$, $\la \db, X_n  \ra=0$ , while
$\zd$
provides the dual to $\db$, i.e.,  $\zd ( \db)=1 \,,$
and $\, \zl_0( \db)=0\,,$ $\za^i (\db)=0\,, i=1,2 \,.$

We will consider weights $\zl \in ({\tt h}\oplus
\dC\db)^* \,$
$$
 (\zl+\zr) (\hb^j) = M^j =  \la  \zl+\zr , \za^j \ra\,,
j=0,1,2\,, $$
 $$
  M^j \in\dC\,,
  \quad M^0+M^1+M^2= k+3   \not = 0\,, \
\eqno(2.6) $$
where
$ \zr (\hb^j) =  \la  \zr , \za^j \ra =1 \,,
j=0,1,2\,, $ and $k= \zl (\kb ) \in \dC\,,$ is the level.
The projection of $\zl$ to  ${\bar {\tt h}}^*$
will be denoted by $\bar \zl$.
We shall not keep track of the
$\zd$ - components $\zl (\db)$, if not necessary, i.e., the
weights will be characterised
by $\{\bar \zl,  \nu=1/ (k+3) \} \ \Leftrightarrow \  \{M^1,M^2; \nu\}.$

For a real root $\beta$ the shifted
action of the Weyl reflection $w_\beta$ on a weight $\lambda$ is
$$
  w_\beta \cdot \lambda =
  w_\beta ( \lambda +\rho) - \rho  =
  \lambda - \langle \lambda+\rho, \beta \rangle \, \beta \,.
\eqno(2.7)
$$
The affine Weyl group $W$ is generated by the three simple reflections
$w_i \equiv w_{\alpha^i}\,$, $i=0,1,2\,.$

A Verma module $\cV_{\lambda}\,$ of highest weight $\lambda$ is
freely
generated by the negative  subalgebra ${\tt n_-} \, $ acting on
the highest weight vector $V_\lambda$
which is a vector of weight $\lambda$, i.e.,
$\hb^j\, V_\lambda =\lambda ( \hb^j) V_\lambda$, $j=0,1,2$,
and $V_\lambda$ is singular,
i.e., it  is annihilated by the raising operators
(it is sufficient if it is annihilated by $\eb_i$, $i=0,1,2$).

Restricting to the case
$\langle\lambda +\rho,\delta \rangle = k+3 \not = 0\,$
the Kac-Kazhdan [15] theorem states that a
Verma module of highest weight $\lambda$ is reducible iff
$$
  \langle \lambda+\rho, \beta \rangle = m \in\dN
  \qquad{\rm for\ some}\ \ \beta\in\Delta_{+,{\rm real}}
\eqno(2.8)$$
and in this case it contains a Verma submodule of highest weight
$ w_\beta \cdot \lambda = \lambda - m \, \beta$.

If $\beta$ of the Kac-Kazhdan theorem is a simple root then one
immediately has an explicit expression for the singular vector
generating the submodule.
Indeed, the following simple fact is true.
If $V_\lambda$ is a singular vector
of weight $\lambda$ such that
$\langle \lambda +\rho, \alpha^j \rangle \in\dN$ then
$$
  V_{w_j\cdot\lambda} \equiv
  (\fb_j)^{\langle \lambda +\rho, \alpha^j \rangle}
  \,V_\lambda \,, \qquad j=0,1,2\,,
\eqno(2.9)$$
is a singular vector. If $i\ne j$ the generators $\eb_i$ annihilate
$ V_{w_j\cdot\lambda} $ because they commute with $\fb_j$. While
for $i=j$ one
has
$ \eb_j\, (\fb_j)^p \, V_\lambda
= p \, ( \langle \lambda +\rho, \alpha^j \rangle - p ) \,
   (\fb_j)^{p-1} \, V_\lambda $
and thus  $V_{w_j \cdot \lambda}$ is annihilated by $\eb_j$ also.

Now we turn to the case of a general positive real root $\beta$.
The Weyl reflection
$w_\beta$ can be written as a product of simple reflections
$w_\beta = w_{i_s}\dots w_{i_3}\,w_{i_2}\,w_{i_1}\,,$
$i_l=0,1,2$ (see below for explicit expressions).
Consider the sequence of vectors
$$
  V_{w_{i_{p}} w_{i_{p-1}}\dots w_{i_1}\cdot\lambda} =
  (\fb_{i_p})^{
         \langle w_{i_{p-1}}\dots w_{i_1}\cdot\lambda + \rho, \alpha^{i_{p}}
 \rangle}
  \, V_{w_{i_{p-1}}\dots w_{i_1}\cdot\lambda} \,,
\eqno(2.10) $$
the last element of this sequence being
the singular vector of weight $w_\beta\cdot\lambda$
$$
  V_{w_\beta\cdot\lambda} = \cP_{\beta;\lambda}  \, V_\lambda
  \quad{\rm with}\quad
  \cP_{\beta;\lambda} \equiv (\fb_{i_s})^{
         \langle w_{i_{s-1}}\dots w_{i_1}\cdot\lambda
         + \rho, \alpha^{i_{s}} \rangle}
  \dots
  (\fb_{i_2})^{\langle w_{i_1}\cdot\lambda + \rho, \alpha^{i_{2}}
\rangle} \,
  (\fb_{i_1})^{\langle \lambda + \rho, \alpha^{i_{1}} \rangle} \,.
\eqno(2.11) $$
If all
$\langle w_{i_{p-1}}\dots w_{i_1}\cdot\lambda
+ \rho, \alpha^{i_{p}} \rangle \in \dN$
then all the vectors of the sequence (2.10) are singular vectors.
In fact a much stronger statement is true. We can relax all these
integrality conditions retaining only the Kac-Kazhdan condition
$\langle \lambda + \rho, \beta \rangle \in \dN$. Even though the
generators are taken to, in general, complex powers and the vectors
(2.10) for $1<p<s$ are only formal expressions, the vector
defined by (2.11)
makes sense as a vector in the Verma module. This is the content
of the Malikov-Feigin-Fuks theorem [16]. In a recent paper [25]
a rigorous definition of complex powers of the step operators is given
and in particular of the vectors in (2.10). In Appendix A we work out
in more details some of the Malikov-Feigin-Fuks vectors.

 For the 6 types of positive real roots $\beta$ of $A^{(1)}_2$
one has the following decomposition of $w_\beta$ into simple
reflections [26]:
$$\eqalignno{
  \beta =\beta_{n,+,1}=
   n\,\delta + \alpha^1
   \qquad & w_{\beta_{n,+,1}} = w_{1(0201)^n}\,,\quad
                                        n = 0,1,2,\dots \,,
&(2.12a)\cr
  \beta_{n,+,2} = n\,\delta + \alpha^2 \qquad & w_{\beta_{n,+,2}}
   = w_{2(0102)^n}\,,\quad
                                        n = 0,1,2,\dots \,,
&(2.12b) \cr
  \beta_{n,+,3} = n\,\delta + \alpha^3 \qquad & w_{\beta_{n,+,3}}
   = w_{121(0121)^n}\,,\
                                        n = 0,1,2,\dots \,,
&(2.12c) \cr
  \beta_{n,-,1}
   = n\,\delta - \alpha^1 \qquad & w_{\beta_{n,-,1}} =
w_{020(1020)^{n-1}}\,,
                                        n = 1,2,3,\dots \,,
&(2.12d) \cr
  \beta_{n,-,2}
   = n\,\delta - \alpha^2 \qquad & w_{\beta_{n,-,2}} =
w_{010(2010)^{n-1}}\,,
                                        n = 1,2,3,\dots \,,
&(2.12e) \cr
  \beta_{n,-,3} = n\,\delta - \alpha^3 \qquad & w_{\beta_{n,-,3}}
   = w_{0(1210)^{n-1}}\,,\quad          n = 1,2,3,\dots \,,
&(2.12f) \cr
}$$
where for short we denote
$ w_{ij(kl\dots)^n\dots}= w_i w_j (w_k w_l \dots)^n \dots\,$.

One can have also  singular vectors arising from
compositions $w_{\beta_s}\dots  w_{\beta_2}w_{\beta_1}$ of Weyl
reflections,
where $\beta_1, \dots ,\beta_s$ are real positive roots,  like
e.g.,
$V_{w_{\beta_2}w_{\beta_1}\cdot\lambda}
= \cP_{\beta_2;\,w_{\beta_1}\cdot \lambda} \, \cP_{\beta_1;
\lambda} \,
V_\lambda$,
if the Kac-Kazhdan condition (8) is satisfied for the root
$\beta_1$
and then  (by the weight $w_{\beta_1}\cdot\lambda$) for the root
$\beta_2$.
Such a vector generates a submodule of weight
$w_{\beta_2}w_{\beta_1}\cdot\lambda$ which is embedded in the
submodule
of weight $w_{\beta_1}\cdot\lambda$.
Here we will not study in detail the Verma module embedding
patterns so
let us give as an illustration only the horizontal singular
vectors. Thus
if both $M^i \in\dN$, $i=1,2$, besides the singular vectors (9)
we have also
$$\eqalign{
  &V_{w_1 w_2 \cdot \lambda} \equiv
  (f^{1}_0)^{M^3}\,(f^{2}_0)^{M^2}\,V_\lambda\,,
   \qquad\qquad V_{w_2 w_1 \cdot \lambda} \equiv
   (f^{2}_0)^{M^3}\, (f^{1}_0)^{M^1}\, V_\lambda\,,
\cr\cr
  & V_{w_2 w_1 w_2 \cdot \lambda} \equiv
   (f^{2}_0)^{M^1}\, (f^{1}_0)^{M^3}\, (f^{2}_0)^{M^2}\,
V_\lambda
= V_{w_1 w_2 w_1 \cdot \lambda} \equiv
   (f^{1}_0)^{M^2}\, (f^{2}_0)^{M^3}\, (f^{1}_0)^{M^1}\,
V_\lambda \,.
\cr}\eqno(2.13)\,,$$
where $M^3=M^1+M^2\,.$
The last vector in (2.13)
corresponds to the Weyl reflection $w_{\beta}=w_1 w_2
w_1=w_2 w_1 w_2\,$ with $\beta=\alpha^1+\alpha^2\,$ and,
being a particular case of (2.11),
it provides also a Malikov-Feigin-Fuks
type singular vector under the
weaker conditions $M^3 \in \dN\,, M^1, M^2$ -- arbitrary.
 \medskip

\bigskip\noindent{\bf 2.2. Chiral algebra fields.}
 \medskip
It is standard in Conformal Field Theory to  consider fields
$$
  A(z) = \sum_{n\in\dZ} A_n \, z^{-n-\Delta_A}
 \eqno(2.14)$$
as generating functions of the modes $A_n$.  The
(anti)commutation relations of the modes are equivalent to the
singular part of the operator product expansion (OPE).  In
particular the KM commutation relations (2.4) can be rewritten as
(the dots indicate the regular terms)
$$
  X(z) \, Y(w) = {k\,\langle X,Y\rangle\over (z-w)^2}
  + {([X,Y])(w)\over z-w} + \dots \
 \eqno(2.15)$$
with the assumption that all the currents $X(z)$ have
$\Delta_X=1$.

The normal product of two fields $(A\,B)(z)$ can be defined as
the zero order term in the expansion of $A$ and $B$. In modes
this is equivalent to
$$
  (A\,B)_n = \sum_{m \le - \Delta_A} A_{m}\, B_{n-m} +
  (-1)^{AB} \sum_{m > - \Delta_A}  B_{n-m} \, A_{m}
 \eqno(2.16)$$
where $(-1)^{AB}$ is $-1$ if both $A$ and $B$ are fermionic and 1
otherwise.  The so defined normal order product is neither
commutative nor associative -- for rules to work with it see
[27].  Since the technique of operator product expansions is well
known and widely used, moreover  a MATHEMATICA code is available
[28],  we will skip all  explicit computations with normal
products needed throughout this work. Let us only write down the
standard formula

$$ [A_m,B_n] =
   \sum_{k\ge 1} {m+\Delta_A -1\choose k-1} (AB)^{(k)}_{m+n} \,,
\eqno(2.17)$$
where as usual $(AB)^{(k)}$ are the coefficients of the OPE $
A(z)B(w)=\sum_{k\le{\rm finite}} {(AB)^{(k)}(w)\over (z-w)^k}$.

The reduced theories are obtained by imposing constraints on
some
of the raising operator currents $e^a(z)$.
In the BRST formalism one needs  for each constraint
a pair of fermionic ghost fields
$b^a$, $c^a$  having operator product expansions
$$
  b^a(z) \, c^b(w) = { \delta_{a,b} \over z-w} + \dots \,.
 \eqno(2.18)$$
and  $\Delta_{b^a}+\Delta_{c^a} =1$. We choose
 $\Delta_{b^a}=0\, , \, \Delta_{c^a}=1$.

%%%%%%%%%%%%%%
%
%   section 3.
%
%%%%%%%%%%%%%%%

\bigskip\noindent{\bf 3. The ${\cal W}$ ALGEBRAS from REDUCTION
of $A_2^{(1)}$.}
\medskip\noindent{\bf 3.1. The ${\cal W}_3$ algebra.}\medskip

In this subsection we will consider the reduction leading to the
${\cal W}_3$ algebra
of Zamolodchikov [21] which is associated to the so called
principal
embedding. The corresponding (classical) constraints are
$$
  e^{1}(z)=e^{2}(z)=1, \qquad e^{3}(z)=0 \,.
\eqno(3.1)$$

Following [8], [11] let us introduce the ``hatted''  generators
$$
   \hat X^a = X^a + f^{a \alpha }_{\beta}\,(b^{\beta}\
c_{\alpha})\,,
\eqno(3.2a)$$
where the summation indices $\alpha, \beta$
correspond to the constrained generators
$e^{\alpha}$,  $\alpha= 1,2,3$. Explicitly
$$\eqalign{
   \hat f^{1} & = f^{1} + (b^2\,c^3)\,, \qquad  \hat
e^1=e^1+(b^3\,c^2)\,,
\cr
   \hat f^{2} & = f^{2} - (b^1\,c^3)\,, \qquad  \hat
e^2=e^2-(b^3\,c^1)\,,
\cr
   \hat f^{3} & = f^{3}\,,\qquad\qquad\qquad    \hat e^3=e^3\,,
\cr
  C_{ij}{ \hat h}^j & = C_{ij}h^j  + (b^ic^i) + (b^3c^3)\,,
\qquad
i=1,2\,,
\cr}\eqno(3.2b)$$
(summation over $j=1,2\,$  assumed).

The BRST charge implementing the  constraints (3.1) is
$$\eqalign{
  Q &= \oint_{C_0} {dz\over 2\pi i} \,
  \left( \sum_{\alpha=1}^3 c^{\alpha} \hat e^\alpha
  +(b^3 (c^1 \,c^2))  - c^1 - c^2 \right)(z)
\cr\cr
  &=\sum_{\alpha=1}^3 (e^{\alpha}\,c^\alpha)_{-1}
      - (b^3 (c^1 \,c^2))_{-1} - c^1_0 - c^2_0 \,.
\cr}\eqno(3.3)$$

The OPE's among the
fields $\{ \hat f, \hat h\}\,$ is the same as among the
corresponding
unhatted ones with the only difference that $k$ is shifted to
$k+3$, i.e.,
$$
  \hat h^i(z)\, \hat h^j (w) = {C^{ij}\over \nu}\, {1\over
(z-w)^2}\,, \qquad
  \nu = {1\over k+3}\,.
\eqno(3.5)$$

Consider the fields
$$\eqalign{
  {1\over\nu}  \Tf (z) &= {1\over 2} \, C_{ij} \, ( \hat h^i\,
\hat h^j)(z)
            + \Big( {1\over\nu}-1 \Big) \, \partial  \hat h^3 (z)
\,,
\cr\cr
  6\aw\, \Wf (z) &=
  2 \, C_{1i}C_{2j} \big(( \hat h^1- \hat h^2)\, \hat h^i\, \hat
h^j \big)(z)
\cr\cr
             &\ +  3 \, \Big( {1\over\nu}-1 \Big) \,
              \left( C_{1i}\, ( \hat h^i\, \partial \hat h^1) -
              C_{2i}\, ( \hat h^i\,\partial \hat h^2) \right)(z)
              + \Big( {1\over\nu}-1\Big)^2 \, \
                \partial^2 ( \hat h^1- \hat h^2)(z)\,
                          \,,
\cr}\eqno(3.6)$$
where $\aw = \epsilon \, \nu^{-3/2}$, $\,\epsilon = \pm 1$; we
shall choose $\epsilon =-1$.

Making the identification
$$
  \partial\phi_1(z) = \sqrt{-\nu} \,\,  \hat h^3(z) \,, \qquad
  \partial\phi_2(z) = \sqrt{-\nu\over 3} \,\, ( \hat h^1 -  \hat
h^2)(z) \,
$$
the fields  $\Tf$, $\Wf$ acquire the form of the free field
realization
of the ${\cal W}_3$ algebra of Fateev and Zamolodchikov  [29].

The  BRST invariant currents are expressed through $\Tf$ and
$\Wf$
according to
$$\eqalign{
  {1\over\nu} \,  T(z) &=
  \hat f^{1}(z)  +  \hat f^{2}(z) +  {1\over\nu} \, \Tf(z) \,,
\cr\cr
  \aw \,  W(z) &=   \left(  \hat f^{3}  +
  {1\over 2} \Big( {1\over\nu} - 1 \Big) \,
      \partial( \hat f^{1}- \hat f^{2}) +
                     C_{2i} \, ( \hat h^i\, \hat f^{1})
                    - C_{1i} \, ( \hat h^i\, \hat f^{2})
\right)(z)
                                  +  \aw \, \Wf(z) \,,
\cr}\eqno(3.7a)$$
or, in modes,
$$\eqalign{
  {1\over\nu}\,L_{n} &=   \hat f^1_{n+1} + \hat f^2_{n+1}  +
  {1\over\nu}\,\Lf_{n}\,=
   \hat f^1_{n+1} + \hat f^2_{n+1}
   + {C_{ij}\over 2}( \hat h^i  \hat h^j)_n +
  \Big({1\over\nu}-1\Big)(\partial  \hat h^3)_n\,,
\cr\cr
   \aw \,W_n  &=    \hat f^{3}_{n+2}  +
  {1\over 2} \Big({1\over\nu}-1\Big)
             \,( \partial\hat f^{1}
                                - \partial\hat f^{2})_{n+1} +
                     C_{2i} \, ( \hat h^i\, \hat f^{1})_{n+1}
                    - C_{1i} \, ( \hat h^i\, \hat f^{2})_{n+1}
                                  + \aw\, \Wf_n \,,
\cr\cr
  \aw \Wf_n &=
  C_{1i}C_{2j} (C_{1l}-C_{2l})
  \,( \hat h^i\,( \hat h^j\, \hat h^l))_n
\cr\cr
            &\ + {1\over 2}\, \Big( {1\over\nu} - 1 \Big) \,
              \left( C_{1i}\, ( \hat h^i\, \partial \hat h^1) -
              C_{2i}\, ( \hat h^i\,\partial \hat h^2) \right)_n
          \, + {1\over 6}\, \Big( {1\over\nu} - 1 \Big)^2
          \, (\partial^2  \hat h^1-\partial^2  \hat h^2)_n \,.
\cr}\eqno(3.7b)$$
These expressions
were computed in [11],
 solving the tic-tac-toe
equations of the BRST double complex.
The algebra of  $T$ and $W$
(which is  identical to that of $\Tf$, $\Wf$) is

$$\eqalign{
  T(z) \, T(w) &=  { c_\nu / 2 \over (z-w)^4}
               + { 2 \, T(w) \over (z-w)^2}
               + { \partial T(w) \over (z-w) } + \dots \, ,
\cr\cr
  T(z) \, W(w) &= { 3\,W(w) \over (z-w)^2 }
                       + { \partial W(w)\over (z-w)} + \dots
\cr\cr
  \beta^2\,  W(z) \, W(w) &=  { c_\nu /3 \over (z-w)^6 }
               + { 2 T(w) \over (z-w)^4 }
               + { \partial T(w) \over (z-w)^3 }
               + {\left( {2\beta^2\over 3}\Lambda
                 + {3\over 10}\partial^2 T \right)(w) \over
(z-w)^2}
\cr\cr
  &\ + {\left( {\beta^2\over 3}
                \partial\Lambda + {1\over 15}\partial^3 T
\right)(w)
        \over (z-w)}   + \dots \,,
\cr}\eqno(3.8)$$
where
$$
  \Lambda = (TT) - {3\over 10} \partial^2 \, T,
  \qquad
  \beta = \sqrt{ 48 \over 22+5c_\nu }
$$
and the conformal anomaly is
$$
  c_\nu = 50 -24\,\Big( \nu + {1\over\nu} \Big) \,.
\eqno(3.9)$$

\medskip\noindent{\bf 3.2.
The Polyakov-Bershadsky algebra.}
\medskip

In the case of $sl(3)$ there is only one $sl(2)$ embedding
besides the principal one. This nonprincipal embedding gives the
Polyakov-Bershadsky algebra ${\cal W}_3^{(2)}$.

The BRST operator is
$$
   Q^{\pb} = \oint_{C_0} {dz\over 2\pi i} \,
   \left( (e^3\,c^3) + (e^2\,c^2) - c^3 \right)(z) =
   (e^3\,c^3)_{-1} + (e^2\,c^2)_{-1} - c^3_0 \,.
\eqno(3.10)$$
This operator corresponds to the classical, first class
constraints
$$
  e^2(z)=0\,,\qquad e^3(z)=1\,.
\eqno(3.11)$$
In the classical consideration the third condition $e^1(z)=0$  is
achieved by a choice of a  gauge fixing $e^1_{{\rm gauged}}=0$.
This is a standard way of dealing with second class constraints
[5].

Following the general scheme of [11] introduce the hatted
quantities
(now $\alpha, \beta$ in (3.2a) run over $2, 3$):
$$\eqalign{
  \hat e^1 &=e^1 + ( b^3 \,c^2 )\,,\qquad \hat e^2=e^2\,,\qquad
\hat e^3=e^3\,,
\cr\noalign{\medskip}
  \hat f^1 &=f^1 + (b^2 \, c^3 )\,,\qquad \hat f^2=f^2\,,\qquad
\hat f^3=f^3\,,
\cr\noalign{\medskip}
  \hat h^1 &= h^1- (b^2 \,c^2)+ (b^3 \,c^3) \,,
\cr\noalign{\medskip}
  \hat h^2 &= h^2+2(b^2\,c^2)+(b^3\,c^3)\,.
\cr} \eqno(3.12)$$

The reduced, BRST invariant  generators [11] are:
$$\eqalign{
       H &= {1\over 3}( \hat h^2- \hat h^1) \,,          \cr
\noalign{\medskip}
       G^+ &= \hat f^1                \,,              \cr
\noalign{\medskip}
       G^- &= \hat f^2 + ( \hat e^1\,\hat h^2)
                + \Big({1\over\nu} - 2\Big)\,\partial \hat e^1
\,,
\cr\noalign{\medskip}
  {1\over\nu} \, T &=  \hat f^3 + (\hat e^1\,\hat f^1)+
{1\over\nu} \, \Tf
\cr
           &=  \hat f^3 + (\hat e^1\,\hat f^1)+
           {1\over 2} C_{ij}\,( \hat h^i\,\hat h^j)
         + \Big({1\over\nu} - 2\Big)\,{\partial \hat h^3 \over 2}
\,.
\cr}\eqno(3.13)$$

One readily checks that they close the ${\cal W}_3^{(2)}$ algebra
(we use interchangeably $k$ and $1/\nu=k+3$):
$$\eqalign{
     H(z)\,H(w) &= {1\over3} {2k+3\over (z-w)^2}
     + \dots \cr
\noalign{\medskip}
     H(z)\,G^{\pm}(w) &= \pm {G^{\pm}(w)\over z-w}
     + \dots \cr
\noalign{\medskip}
     T(z)\,H(w) &= {H(w) \over (z-w)^2} +
      {\partial H\over z-w} + \dots \cr
\noalign{\medskip}
     T(z)\,G^{\pm}(w) &= {3\over 2} {G^{\pm}(w)\over (z-w)^2}
     +{\partial G^{\pm}(w) \over z-w} + \dots       \cr
\noalign{\medskip}
     G^+(z)\,G^-(w) &= {(k+1)(2k+3)\over (z-w)^3} +
     {3(k+1)\over (z-w)^2}\,H(w)         \cr
\noalign{\medskip}
     &\ +{1\over z-w}\left( 3(H^2)(w)-(k+3)\,T(w)+
     {3\over 2}(k+1)\,\partial H(w)\right) + \dots \cr
\noalign{\medskip}
     G^\pm(z)\,G^\pm(w) &= 0 + \dots
\cr}\eqno(3.14)$$
and $T(z)$ closes a Virasoro subalgebra with central charge
$$
   c_{\nu}=25-6\nu-{24\over \nu }\,.
\eqno(3.15)$$

The initial dimensions of $G^{\pm}\,$ inherited
from the Kac -- Moody algebra
are $\triangle_{G^+}=1$, $\triangle_{G^-}=2$.
(These dimensions were used in [20].) Following the
more symmetric standard convention let us use instead half integer modes
arising in an expansion with $\triangle_{G^{\pm}}=3/2\,$ -- which
are also the scale dimensions with respect to $T$. Hence

$$\eqalign{
     H_n &= {1\over 3} (\hat h^2_n - \hat h^1_n)
\cr
\noalign{\medskip}
     G^+_{n- {1\over 2}} &= \hat f^1_n
\cr
\noalign{\medskip}
     G^-_{n- {1\over 2}} &= \hat f^2_{n} + (\hat e^1\,\hat
h^2)_{n-1}
    -\Big({1\over\nu} - 2\Big)\, n\, \hat e^1_{n-1}
           \cr
\noalign{\medskip}
   {1\over\nu}  L_n &= \hat f^3_{n+1} + (\hat e^1\,\hat f^1)_n+
     {1\over 2}C_{ij}\  (\hat h^i\,\hat h^j)_n
    -{1\over 2}\Big({1\over\nu} - 2\Big)\,( n+1)\, \hat h^3_n
\,,
\cr} \eqno(3.16)$$
and one recovers (the Neveu -- Schwarz sector of)
the standard ${\cal W}_3^{(2)} $ mode-algebra.

\bigskip\noindent{\bf 4. QUANTUM GAUGE TRANSFORMATIONS.}
\medskip

We will consider modules  $\,\Omega_{\lambda}$ that are tensor
products of a $A_2^{(1)}\,$ Verma module ${\cal V}_{\lambda} \,$
and a ghost Fock module.  To simplify notation we will avoid
indicating explicitly tensor products, thus assuming that the
highest weight vector $V_\lambda\,$ of $\,{\cal V}_\lambda\,$ is
annihilated also by the positive modes of all $b^i(z)\,$ and the
nonnegative modes of $c^i(z)\,$. Clearly any singular vector in
the module of $A_2^{(1)}\,$ built on $V_{\lambda}\,$ is a
singular vector in  $\,\Omega_{\lambda}\,$ and furthermore we can
use equivalently  the hatted counterparts of the three generating
elements of ${\tt n_-}$ to build these vectors, since
$(b^i\,c^j)_0\,V_{\lambda} = 0\,$ (cf. (3.2a)).  Therefore with
$V_{w\cdot\lambda}$ we denote the singular vectors of both
$\Omega_\lambda$ and ${\cal V}_\lambda$.

The BRST operators (3.3) and (3.10), for the two
reductions respectively, annihilate any singular vector
$V_{w\cdot\lambda}$ (including the highest weight states).
The generators
of the ${\cal W}_3$ and ${\cal W}_3^{(2)}$ algebras have been
expressed in (3.7) and (3.13), respectively,
through the $\widehat sl(3)$ currents and the respective ghosts.
 From these expressions (or better from
their mode versions (3.7b) and (3.16)) it is immediate that
the Kac-Moody singular vectors $V_{w\cdot\lambda}$ are annihilated
by all positive modes of the (respective) ${\cal W}$ algebra.
Also the zero modes of the ${\cal W}$ algebra generators
reduce on such vectors to some polynomial of the zero modes $h^i_0$
of the Cartan currents  and hence to a number, i.e.,
$V_{w\cdot\lambda}$ are eigenvectors of the ${\cal W}$ algebra
zero modes (see (5.3,4), (6.2,3) below for explicit expressions).
Hence in particular we
 can identify the Kac-Moody
highest weight state $V_\lambda$
with the  highest weight state of a  Verma module of the
corresponding $ {\cal W}$ algebra.

For $V_{w\cdot\lambda}$ viewed as Kac-Moody singular vectors
we have explicit expressions
$V_{w\cdot\lambda} = {\cal P} \, V_\lambda$, where ${\cal P}$ are
the Malikov-Feigin-Fuks monomials (or equivalently ordinary
polynomials) of the Kac-Moody generators.
Thus  a Kac-Moody singular vector
will determine a ${\cal W}$ algebra singular vector if it can be
expressed, eventually up to $Q$  exact terms, entirely by
polynomials ${\cal S}$ of ${\cal W}$ algebra generators, i.e.,
${\cal P} \, V_\lambda = ({\cal S} + Q{\rm\ exact\
terms})\,V_\lambda$.
This problem, on the other hand, is far from trivial.

%%%%%%%%%%%%%
%
%     section 4
%
%%%%%%%%%%%%%%%

\medskip\noindent{\bf 4.1. Quantum gauge transformation for
${\cal W}_3$. }
\medskip

Now trying to  ``invert'' the expression (3.7b) for the
negative mode ${\cal W}_3$ generators one gets  in the
simplest case (no summation in $i$)

$$\eqalign{  \Big( M^3-1- {1\over\nu} \Big)
\Big(f^i_0
         &+ (M^i-1) C_{ij}\,  \hat h^j_{-1}\Big)\,V_\lambda
\cr\cr
   &= \left(
(\delta_{i1}-\delta_{i2})\, \aw\, W_{-1}
             + {1\over 2\nu} L_{-1}\,
   \Big( 2 C_{ij}\, M^j -1 - {1\over\nu}\Big)
   \right)\,V_\lambda \,,
   \quad i=1,2\,.
\cr}\eqno(4.1)$$
Comparing with  (2.9) ($i=1,2$) one
finds that for weights $\lambda$ such that $M^1=1$ (or $M^2=1$)
the
l.h.s. of the above equalities become proportional
to  the simplest Kac-Moody singular vectors.
One can expect that there exists in general a map
intertwining  the Kac-Moody and $ {\cal W}_3$ singular vectors.
In principle it could be found using the
Kac-Moody  commutation relations,
systematically ``inverting'' (3.7b) and getting rid at the end of
all Heisenberg subalgebra depending terms. However this
becomes  soon technically rather messy. An idea how
to find such an intertwining map is actually suggested
by the classical Drinfeld-Sokolov approach.

The classical counterparts of the generators (3.7) emerge [1] as
differential  polynomials of the (classical ) fields $f^a(z)$,
$h^i(z)$, and their derivatives, invariant  under the gauge
symmetry generated by the constraints. Their explicit expressions
can be recovered by a  gauge fixing transformation  of the
constrained system --  the so called [3]  ``highest weight''
Drinfeld-Sokolov gauge.  These reduced classical fields can be
quantised by replacing $f^a(z), h^i(z) \rightarrow    \hat
f^a(z),   \hat h^i(z)$, normal ordering and identifying an
intrinsic parameter with $k+2$. This recovers the r.h.s. of (3.7)
with (3.6) taken into account, i.e., the result of [11].  As in
the $\widehat {sl}(2)\,$ case discussed in the Introduction, we
have to look for a proper quantum analogue of this gauge
transformation, generated by the positive subalgebra $ {\tt {\bar
n}_+}$, with gauge parameters depending on   $  \hat h^i(z), \hat
f^a(z)\,$ and their derivatives.  The simplest singular vectors
of $\widehat{sl}(3)$ corresponding to the simple  roots
$\alpha^1, \alpha^2$, are given by  powers of  root vectors
$f_0^i\,$, $\, i=1,2 $, acting on the highest weight state (cf.
(2.9)).  Furthermore the states $(f_0^1)^t\,V_\lambda (= ( \hat
f_0^1)^t\,V_\lambda)$, or $(f_0^2)^t\,V_\lambda (= ( \hat
f_0^2)^t\,V_\lambda)$, are annihilated by $e_0^3$ and $e_0^2$, or
$e_0^3$ and $e_0^1$, respectively.  Thus  we need  not a quantum
analogue of the full original  Drinfeld-Sokolov gauge
transformation, but rather a suitable  ``$sl(2)$''- type
transformation    generated by the simple-root vectors $e_0^1\,$,
or $ e_0^2$, respectively.

So consider the following operators,  generalising
straightforwardly the operator  (1.2),
$$
   {\cal R}^{(i)}(u)
  =  \,\odots \exp  C_{ij}\, \int_0^u du'\     \hat h_{(-)}^j(-u')
   \odots   \,, \quad i=1,2
\,, \qquad
    \hat h_{(-)}^j(u)= \sum_{n=1}^{\infty}u^{n-1} \hat h^j_{-n }\,,
\eqno(4.2)$$
$$   {\cal R}^{(i)}(u)
= \sum_{k=0}\,  {\cal R}_{-k}^{(i)}\,  u^k \,  .$$
The parameter $u$ will be identified with some generators, e.g.,
with  $ \hat e^i_0$.
The  $\odots\ \odots$ indicate that if $u$ and
$C_{ij}  \hat h^j$ do not commute, then in the expansion of the
exponent the
 $u$'s
should come to the right of the modes of $C_{ij}  \hat h^i$. For
a recursive
definition of ${\cal R}^{(i)}$ see  Appendix B.
 \footnote{$^1$}{Note that the coefficients in the expansion of
 $   {\cal R}^{(i)}(u)$
in powers of $u$
 represent elementary Schur polynomials.}
Now we will describe the analogues for $\widehat sl(3)$ of the
basic
properties of the gauge transformations
$$
  {\cal R}^{(i)} \equiv {\cal R}^{(i)}( \hat e^i_0) \qquad
i=1,2\,.
$$
which we discussed in the Introduction.
The choice $u=e^i_0$ or $u= \hat e^i_0\,$, $i=1,2$
is irrelevant for the properties described
below since $(b^3 c^j)_0\,$ annihilate all   states in
$\Omega_{\lambda}$ created  only by Kac-Moody generators, so we
will use equally both.
The first property

\smallskip\item{$\bullet\quad$}
${\cal R}^{(i)}$
{\it leave invariant all Kac-Moody  singular vectors} --

\smallskip\par\noindent
follows from ${\cal R}^{(i)}_0=1$ and the choice $u=\hat e^i_0$.
Denote $V^{(i)}_t\equiv (f^i_0)^t V_\lambda \,$.
For each of the simple roots $\alpha^i$, $i=1,2$,
these vectors  form a   $sl(2)$   Verma module.
The second property says

\smallskip\item{$\bullet\quad$}
${\cal R}^{(i)}$ {\it maps the $sl(2)$ Verma module
corresponding to $\alpha^i$
into the kernel of the BRST operator,}

\smallskip\noindent
i.e.,
$$
  Q\, {\cal R}^{(i)} \, (f^i_0)^t \,V_\lambda =0 \,,
  \quad {\rm any } \  t=0,1,2\dots \,.
\eqno(4.3)$$
The proof is straigtforward: from the OPEs it follows
$$
  [Q,\, C_{ij}\,  \hat h^j_{-n}] = - c^i_{-n}\,,
   \qquad  [Q,\,   \hat e^i_{0}] = 0\,,
  \ \,  i =1,2 \,,
$$
hence  one proves inductively
$$
 [Q,{\cal R}_{-k}^{(i)} ] =  - {\cal R}^{(i)}_{-k+1}\,c^i_{-1}\,,
$$
or,
$$
 [Q,{\cal R}^{(i)} ] = -{\cal R}^{(i)}\, c^i_{-1}\,  \hat e^i_{0}
\,.
\eqno(4.4)$$
Since $V^{(i)}_t\equiv (f_0^i)^t \, V_\lambda$ is annihilated by
all positive mode generators, by $\hat e^3_0$ and $c^a_0$,
$a=1,2,3$, and $\hat e^i_0\,V^{(j)}_t = 0$ for $i\neq j$, we have
$$
  Q\,V^{(i)}_t
  = c^i_{-1}\, \hat e^i_0 \,V^{(i)}_t
$$
which combined with (4.4) proves (4.3).
Finally,

\smallskip\item{$\bullet\quad$}
{\it the gauge transformations intertwine Kac-Moody
 and ${\cal W}$ algebra generators, }

\smallskip\noindent i.e.,
$$\eqalign{
  {\cal R}^{(i)} \, \Big( h^3_0 +2-{1\over\nu} \Big)
  \, f^{i}_0 \, V &= {1\over\nu}
  \sum_{p=1} \left(\left( (\delta_{i1}-\delta_{i2})
\aw\,\nu\,W_{-p}
         - {1\over 2}\Big(1+{1\over\nu}(2C_{ii}-1)\Big)
            \,(\partial L)_{-p} \right.\right.
\cr\cr
  & \hskip 1cm \left.\left.
  - {C_{ii}\over \nu}\, L_{-p} \right) {\cal R}^{(i)}
  + L_{-p}\ {\cal R}^{(i)} \  C_{ij}\, h^j_0 \right)
  \, (-e^{i}_0)^{p-1} \, V \,.
\cr}\eqno(4.5)$$
where $V$ is any vector annihilated by all positive mode generators
and by $c^a_0$, $a=1,2,3$.
Also we have ${\cal R}^{(i)} \,V_t^{(j)}=V_t^{(j)}\,$,
if $i \not = j\,$.

The detailed (rather lengthy) proof of (4.5) is presented in
Appendix B.  In Section 6 this key relation  will be used  to
transform Kac-Moody singular vectors into  $ {\cal W}_3\,$
algebra ones.  The second property (4.3) suggests that the states
$\{ {\cal R}^{(i)} \,\,(f^i_0)^t \,V_\lambda , t=0,1,2,\dots \}\,
$ are actually  elements in the universal enveloping (negative
mode) subalgebra of $ {\cal W}_3$.

Note that, as in the $sl(2)$ case [17],
one can  reformulate (4.5)  identifying the parameter $u$
with the  raising operators $t^{+,i}$, $i=1,2$, of an
auxiliary $sl(3)$ algebra $\{t^a\}\,,$ instead of $ \hat e^i_0$.

\bigskip\noindent{\bf 4.2.
Quantum gauge transformation for the
 Polyakov-Bershadsky algebra.}
\medskip

Now we turn to ``inverting'' (3.16), i.e., expressing the
Kac-Moody generators in terms of ${\cal W}_3^{(2)}$
generators, at least in the Malikov-Feigin-Fuks monomials.
Again the key role is played by the projections
${\cal R}^{(i)}$ along simple
root directions $\alpha^i$, $i=1,2$, of
the respective ``quantum gauge transformation''. The
operators ${\cal R}$ have properties analogous to the
ones discussed in the introduction and in the previous
subsection, only now things are much simpler.
Of course, the remarks at the begining of
subsection 4.1 apply in the present case also.

Since $G^+_{-{1\over 2}} =\hat f^1_0\,$ the corresponding gauge
transformation ${\cal R}^{(1)}\,$ is
(as in the classical case) just an identity.
For the second (horizontal) direction
introduce the operator:
$$
  {\cal R}^{(2)}(u) =  \odots \exp \hat e^1_{-1}\,u \odots
  \equiv \sum_{n=0} {1\over n!} (\hat e^1_{-1})^n\, u^n\,.
\eqno(4.6)$$
Choosing the generator of the gauge transformation to be
$u=e^2_0$ (the relevant properties of ${\cal R}$ are
unaltered if we choose $u = \hat e^2_0$) we set
$$
  {\cal R}^{(2)}\equiv {\cal R}^{(2)}(e^2_0)\,.
$$

The first property of ${\cal R}$ is obvious.
As before denote $V^{(2)}_t = (f^2_0)^t\,V_\lambda$.
{}From $[\hat e^1_{-1},\,Q^{\pb}] = c^2_{-1}$,
$[\hat e^1_{-1}, c^2_{-1}]=0$, and
$[e^2_0,\,Q^{\pb}] = 0$
we get
$$
  [{\cal R}^{(2)},\,Q^{\pb}] = {\cal R}^{(2)}c^2_{-1}e^2_0\,,
$$
which combined with
$Q^{\pb}V^{(2)}_t = c^2_{-1}e^2_0 V^{(2)}_t$
gives the second property of the gauge transformation:
$$
  Q^{\pb}\, {\cal R}^{(2)} \, V^{(2)}_t = 0\,.
\eqno(4.7)$$

Finally the gauge transformation ${\cal R}^{(2)}$
has the intertwining property
$$
  {\cal R}^{(2)}\, f^2_0 \,V^{(2)}_t
  = G^-_{-{1\over 2}}\,{\cal R}^{(2)}\, V^{(2)}_t\,.
\eqno(4.8)$$
The proof is straightforward. Notice that
from the definition (3.16) of $G^-_{-{1\over 2}}$
and the commutation relations we get
$ G^-_{-{1\over 2}}\,{\cal R}^{(2)}V^{(2)}_t =
(f^2_0+e^1_{-1}h^2_0)\,{\cal R}^{(2)}V^{(2)}_t$.
Using the Kac-Moody commutation relations we have furthermore
$(f^2_0+e^1_{-1}h^2_0)\,{\cal R}^{(2)} =
{\cal R}^{(2)}\,f^2_0\,$ and (4.8) follows.

Repeating (4.8) until ${\cal R}^{(2)}\,$ reaches $V_{\lambda}\,$
we obtain
$${\cal R}^{(2)}\,  (f^2_0)^{t+1} V_{\lambda} = (G^-_{-{1\over
2}})^{t+1}\, V_{\lambda}
\,.\eqno(4.9)$$

%%%%%%%%%%%%%%
%
%%%                 section 5
%
%%%%%%%%%%%%%%%%%%

\bigskip\noindent{\bf 5.
SINGULAR VECTORS of ${\cal W}_3^{(2)}$ VERMA MODULES.}
\medskip

Having the quantum gauge transformations we turn now to
the description of the ${\cal W}$ algebras singular vectors.
In this section
we will deal with the simpler case of ${\cal W}_3^{(2)}$. The
strategy is the same as in the Virasoro case [17] -- for a
general
Malikov-Feigin-Fuks monomial use the quantum constraint
$$
  {\bf f}_0 (\equiv e^3_{-1}) = 1 + \{Q^{\pb}, b^3_0\}\,
\eqno(5.1)$$
in the direction of the affine simple root $\alpha^0$,
while in the directions of the non-affine simple roots
use the quantum gauge transformations.

The remarks and notational convention from the begining
of section 4 hold. More explicitly we have
$$
  B_n\,V_{w\cdot\lambda} = 0\, \quad {\rm for }
  \quad B_n\, = L_n\,, H_n\,, \ \  {\rm or }
  \quad  G_{n - {1\over 2}}^{\pm}
  \,, \quad \,n  \in \dN\,,
\eqno(5.2)$$
i.e., the positive modes of the ${\cal W}_3^{(2)}$
generators (see (3.16))
annihilate any Kac-Moody singular vector $V_{w\cdot\lambda}$.
These singular vectors are also eigenvectors of the
zero modes
$$
  L_0\,V_{w\cdot\lambda} =
  h_{w\cdot\lambda}^{{\pb}}\,V_{w\cdot\lambda} \,,
\qquad
  H_0 \,V_{w\cdot\lambda} =
  q_{w\cdot\lambda} \, V_{w\cdot\lambda}  \,,
\eqno(5.3)$$
with
$$
  h^{\pb}_\lambda
  = {\nu\over 2} \langle\ol,\zl-\kappa^{\pb}\rho\rangle \,,
\qquad
  q_\lambda =  (C_{2j}-C_{1j})\, \langle\ol,\alpha^j\rangle \,,
  \qquad\qquad \kappa^{\pb} = {1\over\nu} - 2 \,.
\eqno(5.4)$$

In particular we can identify
$V_\lambda$ with the highest weight state
$|q_\lambda, h^{\pb}_\lambda\rangle\,$
of a ${\cal W}^{(2)}_3$ Verma module.
In fact the correspondence between Kac-Moody singular
vectors and ${\cal W}^{(2)}_3$ ones is not 1 to 1.
 Indeed, the ${\cal W}_3^{(2)}$ algebra weights (5.4)
are invariant under the shifted action of $w_0$, i.e.,
$$
  h^{\pb}_\lambda =  h^{\pb}_{w_0 \cdot \lambda}\,,
  \qquad q_\lambda=q_{w_0 \cdot \lambda}\,,
\eqno(5.5)$$
or, equivalently -- under
a ${1\over 2}\kappa^{\pb}$-shifted
action on the projected weights $\ol$ of
the reflection in the $\alpha^3$ direction,
$\ol'-{1\over 2}\kappa^{\pb}\bar{\rho}
= w_{\alpha^3}(\ol-{1\over 2}\kappa^{\pb}\bar{\rho})\,$,
$\ol' =  \overline{ w_0\cdot \lambda} \,$.
Since a Verma module is determined uniquely by the
highest weights we have to identify
$|q_\lambda, h^{\pb}_\lambda\rangle\,$
with the pair of Kac-Moody
highest weight vectors
$V_\lambda$ and $V_{w_0\cdot\lambda}$.

Now let $\lambda$ be such that it satisfies the Kac-Kazhdan
condition (2.8) with respect to the affine root $\alpha_0$, i.e.
$\langle\lambda+\rho,\alpha^0\rangle=m\in\dN$. The The Kac-Moody
singular vector in (2.9) corresponding to  $\za^0$, does not
produce nontrivial singular vectors in the ${\cal W}_3^{(2)}$
Verma module.  The reason is that $  V_{w_0 \cdot \lambda}$ is
cohomologically equivalent to the highest weight state $V_{
\lambda}\,,$ i.e.,

$$
  V_{w_0 \cdot \lambda}
  =  (1 + Q^{\pb} \cdots) V_{ \lambda}
  \simeq V_{\lambda}
\eqno(5.6)$$
(With $\simeq$ we denote equality up to $Q$-exact terms and note
that since (5.2) holds for both  $V_{\lambda}$ and $V_{w_0 \cdot
\lambda}$ it holds for the $Q$ - exact terms as well.) This is
consistent with the symmetry (5.5) of the zero modes eigenvalues.

The invariance subgroup $\overline W^{(\eta)}=\{1,w_0\}$ of the
Weyl group $W$ is directly connected with the defining vector of
the $sl(2)$ embedding, in the case under consideration this is
$\eta=\alpha^3/2$, and the (nontrivial) constraint, here (5.1).
In the ${\cal W}_3^{(2)}$ case one is constraining to 1 the
current of the $\alpha^3$ root vector, or in modes
$e^3_{-1}\simeq 1$ and note that $e^3_{-1}$ is the root vector
of $-(\delta - \alpha^3) =-\alpha_0\,.$

Now let us turn to the simplest Kac-Moody singular vectors
corresponding to the simple reflections $w_i \,, i=1,2\,$. Let
$M^i\equiv\langle\lambda+\rho,\alpha^i\rangle\in\dN$ for $i=1$ or
2, then the corresponding  KM singular vectors are given in
(2.9).  Using (3.16), the iterated relation (4.9), and the fact
that ${\cal R}$ keeps singular vectors invariant,  we obtain

$$
  V_{w_1 \cdot \lambda} =  (f^1_0)^{M^1} \, V_\lambda =
  (G^+_{-{1\over 2}})^{M^1 }\,V_\lambda \,,
  \eqno(5.7a)$$ or,
$$  V_{w_2 \cdot \lambda}  = (f^2_0)^{M^2
  } \, V_\lambda =
  (G^-_{-{1\over 2}})^{M^2 }\,V_\lambda \,.
\eqno(5.7b)$$

The other horizontal KM singular vectors (2.13) are represented
similarly by compositions of the operators in the r.h.s. of
(5.7), e.g., starting from the first vector in (2.13) and using
the properties of the gauge transformation, equivalently (5.7),
the following chain leads us to an expression for $V_{w_1
w_2\cdot\lambda}$ entirely in terms of ${\cal W}_3^{(2)}$
algebra generators:
$$   V_{w_1 w_2 \cdot \lambda}
  =  (f^{1}_0)^{M^3}\, V_{w_2 \cdot \lambda}
  = (G^+_{-{1\over 2}})^{M^3} \, V_{w_2 \cdot \lambda}
  = (G^+_{-{1\over 2}})^{M^3}\,(G^-_{-{1\over 2}})^{M^2}
     \,V_\lambda \,.
\eqno(5.8)$$

Next we consider a general Kac-Moody singular vector
represented by a Malikov-Feigin-Fuks  monomial
$V_{w_\beta\cdot\lambda}
={\cal P}_{\beta;\lambda}\,V_\lambda$
(see Appendix A for explicit expressions).
The procedure of expressing ${\cal P}_{\beta;\lambda}$
in terms of ${\cal W}_3^{(2)}$ algebra generators,
modulo $Q$ exact terms, consists of moving from left
to right and exploiting the properties of the BRST
operator $Q^\pb$ and the gauge operator ${\cal R}$.
How ${\cal R}$ transforms the factors corresponding
to simple non-affine reflections is clear from the
above example, so consider a factor corresponding to
$w_0$. Let $w_\beta=w'\,w_0\,w''$ for some
$w',w''\in W$ and assume all factors corresponding to
$w'$ have been expressed in terms of ${\cal W}_3^{(2)}$
algebra generators, more precisely --
$G^\pm_{-{1\over 2}}$'s. {}From the quantum constraint (5.1),
or equivalently from (5.6), we have
$$
  V_{w_\beta\cdot\lambda} \simeq
  \underbrace{\cdots\cdots}_{ {\cal W}\ {\rm generators} }
  \,(e^3_{-1})^{\langle w''\cdot\lambda+\rho,\alpha^0\rangle}
  \,V_{w''\cdot\lambda}
  =\underbrace{\cdots\cdots}_{ {\cal W}\ {\rm generators} }
  \,(1 + Q^\pb A)
  \,V_{w''\cdot\lambda}\,,
\eqno(5.9)$$
for some $A$.
Because $Q^\pb$ commutes with the ${\cal W}_3^{(2)}$ algebra
generators it can be pulled to the left
producing a $Q$ exact contribution. The end result is very simple
--

\medskip\noindent{}\ {}  \quad
{\it in a Malikov-Feigin-Fuks monomial substitute
     $\{ f^1_0, f^2_0, e^3_{-1}\}$ with
     $\{ G^+_{-{1\over 2}} , G^-_{-{1\over 2}} , 1 \}$
}\hfill (5.10)\medskip

As an illustration consider $\lambda$ such that
$M^1= m_1 -{m_1'-1\over \nu}\,,$ with  $m_1,m_1' \in \dN$
and take the Kac-Kazhdan root
$\beta =(m_1'-1) \delta +\alpha^1$. Then starting from the
explicit expression for the MFF vector written in  (A.1) of
Appendix A one obtains
$$
  V_{w_{\beta} \cdot \lambda}   \simeq
  (G^+_{-{1 \over 2}})^{m_1+ {m_1'-1 \over \nu}}
  \, (G^-_{-{1\over 2}})^{m_1+ {m_1'-2 \over \nu}}  \, \dots
  (G^+_{-{1 \over 2}})^{m_1- {m_1'-3 \over \nu}} \,
  (G^-_{-{1 \over 2}})^{m_1- {m_1'-2 \over \nu}} \,
  (G^+_{-{1 \over 2}})^{M^1} \,V_\lambda \,.
\eqno(5.11)$$
Using (5.2-4) one sees that (5.11) provides a singular vector in
the ${\cal W}_3^{(2)}$ Verma module built on
$|q_{\lambda^{}}\,,h^{\pb}_{\lambda^{}}\rangle
\equiv  V_\lambda\,$ (whose
$L_0$ level is  $m_1 (2m_1'-1)/2$).

The symmetry (5.5) under the two element group $\{1,w_0\}$
implies that the vector (5.11) can be alternatively obtained
starting from the KM singular vector $  V_{w_\beta' w_0
\cdot\lambda} = V_{ w_0 w_\beta \cdot\lambda} $ in the KM Verma
module $\Omega_{w_0 \cdot \lambda}$, where $\beta' = m_1' \delta
- \alpha^2$.  Indeed, from the explicit expressions for ${\cal
P}_{\beta; \lambda}\,$ (compare  (A.1) and (A.3) - with  ``1''
and ``2'' interchanged), the rule (5.10) and (5.5) we obtain

$$
  V_{w_\beta\cdot\lambda} \simeq
  V_{w_{\beta'}\cdot\lambda'}\,,
  \quad{\rm with}\quad
 \beta'=w_0(\beta),\ \lambda'=w_0\cdot\lambda \,,
\eqno(5.12)$$
which extends (5.6).

The other two series, corresponding to the roots
$\zb =(m_2'-1) \zd +\za^2\,,$ $m_2' \in \dN\,$, and $\zb
=(m_3'-2) \zd +\za^3\,$ $m_3'-1 \in \dN\,$, for which
$M^2= m_2 -{m_2'-1\over \nu}\,,$ $m_2 \in \dN\,,$ and
$M^3= m_3 -{m_3'-2\over
\nu}\,,$ $m_3 \in \dN\,,$ respectively, are obtained in exactly
the same way, starting from the explicit expressions (A.3), (A.4)
for the Malikov-Feigin-Fuks vectors.
As in (5.12) these ${\cal W}_3^{(2)}$ singular  vectors admit
alternative derivations using that
$w_0( \zb_{n,+,2})= \zb_{n+1,-,1}$ and
 $w_0( \zb_{n,+,3})= \zb_{n+2,-,3}$ respectively.

The question arises to give meaning to formal expressions like
(5.11) and proving rigorously that the r.h.s. is indeed a  ${\cal
W}_3^{(2)}$ singular vector.  Rather than repeating the
argumentation from [16] we will describe an algorithm which
transforms expressions like (5.11) into ordinary polynomials of
${\cal W}_3^{(2)}$ algebra generators acting on the h.w. state.
First
note that the commutation relation
$$\eqalign{
  G^-_{-p-{1\over 2}} \, (G^+_{-{1\over 2}} )^{s} =
  (G^+_{-{1\over 2}} )^{s} \, G^-_{-p-{1\over 2}}
  &- s\,(G^+_{-{1\over 2}} )^{s-1}
  \,\Big(3(HH) + {3\over 2}\,\partial H - (k+3)\,T\Big)_{-p-1}
\cr\cr
  &- s(s-1)\,(G^+_{-{1\over 2}} )^{s-2}\,
  \Big(3(G^+H) + (k+3)\,\partial G^+\Big)_{-p-{3\over 2}}
\cr\cr
  &- s(s-1)(s-2)\,(G^+_{-{1\over 2}} )^{s-3}\,(G^+G^+)_{-p-2}
\cr}\eqno(5.13)$$
(and the analogous one with $G^\pm$ reverted)
can be analytically continued in $s$. (The other necessary
commutation relations do not involve non-linearities and
thus are simpler and are left to the interested reader.)
To obtain (5.13) one has to use (2.17).

Next, transforming a Malikov-Feigin-Fuks monomial by the
procedure (5.10), results in a monomial having
a property analogous to (A.5), i.e., in the middle there
is a generator to an integer power, it is surrounded by
a generator with powers adding to an integer and so on.
Thus the algorithm is (in complete analogy with the one,
described at the end of Appendix A,
for Malikov-Feigin-Fuks vectors) --
{\it start from the middle and using
relations like (5.13) move outwards}.

Let us see it explicitly in the case $m_1=1$, $m_1'=2$, i.e.
$$
  \beta=\delta +\alpha^1,\qquad M^1=1-{1\over\nu} \,.
\eqno(5.14)$$
Then (5.11) reduces to
$$
  V_{w_\beta\cdot\lambda}\simeq
  (G^+_{-{1\over 2}})^{1+{1\over\nu}}\,
  G^-_{-{1\over 2}}\,(G^+_{-{1\over2}})^{1-{1\over\nu}}
  \,V_\lambda\,.
\eqno(5.15)$$
The general formula (5.13) simplifies when applied on a
singular vector, in particular
$$\eqalign{
  G^-_{-{1\over 2}}(G^+_{-{1\over 2}})^s\,V_\lambda =
  (G^+_{-{1\over 2}})^{s-2}
  \Big[ s(k & + 3)\,G^+_{-{1\over 2}}\,L_{-1}
   - 3s(2q_\lambda+s-1)\,G^+_{-{1\over 2}}\,H_{-1}
\cr\cr
  & - s(s-1)(3q_\lambda+2s-1)\,G^+_{-{3\over 2}} +
      (G^+_{-{1\over 2}})^2\,G^-_{-{1\over 2}}\Big]
  \,V_\lambda \,,
\cr}\eqno(5.16)$$
and setting $s=1-{1\over\nu}=-k-2\,,$
for the r.h.s. of (5.15), we get
$$\eqalign{
  V_{w_\beta\cdot\lambda}\simeq
  \Big[ - (k+3)(k & + 2)\,G^+_{-{1\over 2}}\,L_{-1}
   + 3(k+2)(2q_\lambda -k-3 )\,G^+_{-{1\over 2}}\,H_{-1}
\cr\cr
  & - (k+3)(k+2)(3q_\lambda - 2k - 5)\,G^+_{-{3\over 2}}
  + (G^+_{-{1\over 2}})^2\,G^-_{-{1\over 2}}\Big]\,V_\lambda\,.
\cr}\eqno(5.17)$$
One can independently check (5.2), i.e., that the above
expression
is indeed a ${\cal W}_3^{(2)}$ algebra singular vector.
For example a rather lengthy computation gives
$$\eqalign{
  G^-_{1\over 2}\,V_{w_\beta\cdot\lambda} & =
  (M^1-1+{1\over\nu} )(M^2-{2\over \nu }+1)
  \,\big[ -(k+3)(k+2) \, L_{-1}
\cr\cr
  & + [2(k+2) (M^2-M^1) - 3(k+1)(k+4)]H_{-1}+
  2G^+_{-{1\over 2}}\,G^-_{-{1\over 2}}\big]\, V_\lambda\,.
\cr}\eqno(5.18)$$
We remark that the two factors
$(M^1-1+{1\over\nu} )(M^2-{2\over \nu } +1)$
appear in all the r.h.s.'s of
$B_n\,V_{w_\beta\cdot\lambda},\quad n \in \dN $.
The first factor reflects
the Kac-Kazhdan condition (5.14), while the second
one comes from the Kac-Kazhdan condition $M^2={2\over \nu }-1$
related to the highest weight $w_0\cdot\lambda$ and
$\beta'=w_0(\beta)= 2 \,\delta  -\alpha^2$
in agreement with (5.12).

Having the singular vectors one can analyse the reducibility
structure of ${\cal W}_3^{(2)}$ Verma modules for different
values of the central charge $c_{\nu} $. It is clear that it will
be very similar to the structure of the KM Verma modules, modulo
the identifications accounting for the symmetry (5.5).  Hence
also the  ${\cal W}_3^{(2)}$ characters can be found following
the standard procedures, or directly from the $A_2^{(1)}$
characters, as done in [13] for the case of the principal
embeddings.

%%%%%%%%%%%%%%%%%
%
%      section 6.
%
%%%%%%%%%%%%%%%%%

\bigskip\noindent{\bf 6.
SINGULAR VECTORS of ${\cal W}_3$ VERMA MODULES.}
\medskip\noindent{\bf 6.1. The basic ${\cal W}_3$ algebra
singular vectors.}\medskip

While the derivation of the $\W_3^{(2)}\,$ algebra singular
vectors is appparently  quite analogous to the one in the
$\widehat {sl}(2)$ case [17], the case $\W_3\,$ is more involved.
We begin with describing how the quantum gauge transformations of
section 4.1. can be used to recover the simplest series of ${\cal
W}_3\,$ singular vectors obtained first by Bowcock and Watts
[23].

Let   $V_{w \cdot \zl}\,$ be  some $\hs\,$  singular vector
in $\,\Omega_{\zl}\,$ of weight $w \cdot \zl\,$.
As already discussed in the beginning of section 4 all
Kac-Moody singular vectors are annihilated by the positive modes
of the
reduced $\W\,$ generators (3.7b), i.e.,
$$
  L_n \, V_{w \cdot \zl} =  \Lf_n \, V_{w \cdot \zl}=0
 \,,\qquad
  W_n \, V_{w \cdot \zl} = \Wf_n \, V_{w \cdot \zl}=0
 \,, \quad n \in \dN \,.\eqno(6.1)$$
 For the zero modes one obtains
$$
  L_0 \, V_{w \cdot \zl} =  \Lf_0 \, V_{w \cdot \zl}=
    h^{(2)}_{w \cdot \zl}   \, V_{w \cdot \zl} \,,\qquad
  W_0 \, V_{w \cdot \zl} = \Wf_0 \, V_{w \cdot \zl}
  = h^{(3)}_{w \cdot \zl} \, V_{w \cdot \zl} \,,
\eqno(6.2)$$
where
$$\eqalign{
  h^{(2)}_{\zl} &=
   {\nu \over 2}C_{ij}\, \zl (h_0^i)\,(\zl  -2\zk \zr)(h_0^j)
\,  =
\,   {\nu\over 2} \la\ol,\zl-2\zk\rho\ra \,,
\qquad \zk = \kk - 1 \,,
\cr
{}
\cr
\aw   h^{(3)}_\zl &= C_{1i}\, C_{2j}\, (C_{1l}-C_{2l})\,
  \la\zl-\zk\rho,\za^i\ra \,
  \la\zl-\zk\rho,\za^j\ra \, \la\zl,\za^l\ra \,.\cr}
\eqno(6.3)$$

We shall start with the simplest subseries of Kac-Moody singular
vectors given in (2.9) for $j=1,2$. We will describe how the key
relation (4.5) transforms a Kac-Moody singular vector of this
type into a $\W_3\,$ algebra one. Take $V=\Vi_{t-1}\equiv
(f^i_0)^{t-1}\, \V$. Then (4.5) can be rewritten as
$$
  \Ri\,\, \Vi_t =
  \sum_{p=1}\, \cL^{(i)}_{t,t-p}\,\,\Ri\,\,\Vi_{t-p}
\eqno(6.4)$$
where we have denoted
$$\eqalign{
  \cL^{(i)}_{t,t-p}\, &=\, {\prod_{l=1}^{p-1}
  (l-t)(M^i+l-t)\over
                     \nu \,  (M^3  - \kk - t)}\, \cdot
\cr\cr
  &\ \cdot\left(
  (\zd_{i1}-\zd_{i2}) \aw \, \nu \,W_{-p}
   + \Big( C_{ij}\,M^j -1- t + p -
  {2\over 3\nu}\Big)  L_{-p}
  - \Big( {1\over 2} + {1\over 6\nu} \Big) (\p L)_{-p} \right)
\,.
\cr}\eqno(6.5)$$
Iterating (6.5) until we reach the singular vector $\V$
 we  obtain
$$
  \Ri\,(f^i_0)^t\, \V =
   \cO^{(i;t)}_{\zl} \, \V
\eqno(6.6)$$
where according to (4.3) the r.h.s. of (6.6) is in the kernel of
the operator $Q$. Explicitly it reads
$$
   \cO^{(i;t)}_{\zl} \equiv
      \cO^{(i;t)}_{(M^1,M^2;\nu)} =
              \sum_{k=1}^t \sum_{\{p_a\}_{a=1}^{k-1}}
  \cL^{(i)}_{t,p_{k-1}} \,  \cL^{(i)}_{p_{k-1},p_{k-2}}
   \dots \, \cL^{(i)}_{p_1,0}
\,,\eqno(6.7)$$
where the second sum is over all $\{t>p_{k-1}>\dots>p_1>0\}$.

Denote for short $\cO^{(i)}_{\zl} =\cO^{(i;M^i)}_{\zl} \,$.
Using  (6.6), the property that $\Ri$ leaves singular vectors
invariant and (6.1,2), which in particular implies that we can
identify $V_{\zl}\,$ with the highest weight state $|h^{(2)}_{
\zl}\,, h^{(3)}_{ \zl}\ra\,$ of a ${\cal W}_3\,$ algebra Verma
module, we obtain:
\medskip

\par\noindent{\it The
 Kac-Moody singular vectors} (2.9) ($i=1,2$)
{\it are equal to $\W_3$ algebra singular vectors}
$$
V_{w_i \cdot \zl} =  (f^{i}_0)^{M^i}\, \V =
  \cO^{(i)}_{\zl}   \, \V\,.
\eqno(6.8)$$

\par\noindent{\it
Furthermore, this is true for  all horizontal Kac-Moody
 singular vectors} (2.13),
$$\eqalign{
V_{w_1 w_2 \cdot \zl}=  (f^{1}_0)^{M^3}\, (f^{2}_0)^{M^2}\, \V
&=
 \cO^{(1)}_{w_2 \cdot \zl} \, V_{w_2 \cdot \zl}
  = \cO^{(1)}_{w_2 \cdot \zl} \, \cO^{(2)}_{\zl}  \, \V \,,
\cr\cr
V_{w_2 w_1 \cdot \zl}=  (f^{2}_0)^{M^3}\, (f^{1}_0)^{M^1}\, \V
&=
\cO^{(2)}_{w_1 \cdot \zl} \, \cO^{(1)}_{\zl}  \, \V \,,
\cr\cr
V_{w_2 w_1 w_2 \cdot \zl}=
  (f^{2}_0)^{M^1}\, (f^{1}_0)^{M^3}\, (f^{2}_0)^{M^2}\, \V &=
  \cO^{(2)}_{w_1 w_2 \cdot \zl} \, \cO^{(1)}_{w_2 \cdot \zl} \,
  \cO^{(2)}_{ \zl} \, \V \,.
\cr}\eqno(6.9)$$

Let us recall that in the first and the second equalities in
(6.9) both $M^1$ and $M^2$ are nonnegative integers, while in
the third one it is sufficient that $M^3=M^1+M^2$ is a
nonnegative integer.  Thus already here one encounters the
need for a proper analytic continuation of the basic vectors in
(6.8).

Thus we recover the result of [23] obtained by a different method
-- namely, by a generalisation of the fusion method of [24]
exploited in the Virasoro case.  Note that in view of   (6.1,2)
the identification (6.8), (6.9), makes  the proof of singularity
of the $\W_3$  states in  the right hand sides straightforward
in our approach.

The states (6.6,7) for a given $i$ and any $t=0,1,2,\dots,$
provide the basis for a matrix reformulation,
in analogy with the one in [24], of the $\W_3$
singular vectors of the type considered up to now.

\medskip
\medskip\noindent{\bf 6.2. The general ${\cal W}_3$ algebra
singular vectors.}\medskip

We turn to a systematic description of the general
${\cal W}_3\,$
singular vectors following the strategy outlined   in [20].

The Kac-Moody singular vectors in (2.9),
 corresponding to the affine root $\za^0$,
produce now $Q$-exact terms, i.e.,
$$
 V_{w_0 \cdot \zl}
  \simeq 0  \,,
\eqno(6.10)$$
because of the quantum constraint
$$\fb_0 =  \{ Q, b^3_0 \}\eqno(6.11)$$
(recall that $\fb_0 = \he^3_{-1}$ ) and the properties of the BRST
charge $Q$.

Thus we cannot exploit exactly the same  mechanism as in the case
of the $\W_3^{(2)}$ Verma modules in order to construct the
general $\W_3$ singular vectors. On the other hand
the modes of the currents which are now constrained to 1, i.e.,

 $$ \hat e^i_{-1} = 1 + \{Q,b^i_0\}\,, \quad i=1,2 \,,
 \eqno(6.12)$$
do not appear explicitly in the Malikov-Feigin-Fuks monomials.
Yet one notices that the affine reflections $w_{\beta}$ in (2.12),
 underlying the construction of the MFF vectors, contain in their
decompositions elements of the type $w_0 w_1 w_0=
w_{\delta-\alpha^2}$ and  $w_0 w_2 w_0=w_{\delta-\alpha^1}$.
The root vectors corresponding to these reflections are exactly
the modes constrained in (6.12). Furthermore these elements of
the affine Weyl group generate the finite subgroup
$\overline{W}^{(\rho)}$, which keeps invariant (under the shifted
action on $\zl$) the zero mode eigenvalues (6.3), i.e.,

$$
  h^{(2)}_\zl = h^{(2)}_{w\cdot\zl}\,, \quad h^{(3)}_\zl =
  h^{(3)}_{w\cdot\zl}
  \,, \quad w \in
\overline{W}^{(\rho)}\,.\eqno(6.13)$$
This is equivalent to the well known invariance under a
$\zk$-shifted action $\ol'-\zk\bar{\rho} = w(\ol-\zk\bar{\rho})$
of the finite Weyl group  $\overline{W}$ on the projected
weights.

Thus   we can identify the highest weight state $| h^{(2)}_{
\zl}\,, h^{(3)}_{ \zl}\ra\,$ of a ${\cal W}_3\,$ algebra Verma
module with  the set of Kac-Moody highest weight states
$\{V_{w\cdot\zl};\,  w \in \overline{W}^{(\rho)} \}$.

Explicitly the group $\overline{W}^{(\rho)}$  acts on the
weights according to

$$ \overline{w_{010}\cdot \zl +\rho}=\left(M^3-\kk\,,
\, -M^2 +
{2\over \nu} \right)\,,\eqno(6.14a)$$
$$  \overline{w_{020}\cdot \zl +\rho}=\left( -M^1 +
{2\over \nu}\,, \,M^3-\kk \right)\,,\eqno(6.14b)$$
$$ \overline{w_{0210}\cdot \zl +\rho}=\left( -M^3 +
{3\over \nu}\,, \,M^1 \right)\,,\eqno(6.14b)$$
$$ \overline{w_{0120}\cdot \zl +\rho}=\left(M^2\,, \, -M^3 +
{3\over \nu} \right)\,,\eqno(6.14d)$$
$$ \overline{w_{01210}\cdot \zl +\rho}=\left(
{2\over \nu}-M^2\,, \,{2\over \nu}-M^1 \right)\,.\eqno(6.14e)$$

Now let us start with the Kac-Moody singular vectors
 $V_{w_\beta\cdot\lambda}\, $ for
$\beta=\delta-\alpha^1\,, $ or $\beta=
\delta-\alpha^2\,$, which occur in
the modules $\Omega_{\lambda}\,$ when $\lambda$ satisfy
the Kac-Kazhdan
condition (2.8), i.e., for $M^1= -m_1+{1\over \nu}\,, $  $m_1\in
\dN\,,$ or $M^2=
-m_2+{1\over \nu}\,, $  $m_2\in \dN\,,$ respectively.
The singular vectors
$V_{w_{\delta-\alpha^i}\cdot \lambda} $ are
cohomologically equivalent to the
corresponding  highest weight states  $V_{\lambda}\,$, i.e.,
$$
  V_{w_{\delta-\alpha^i}\cdot \lambda}\, \simeq A_i\,
  (\hat e^i_{-1})^{m_i} \,
  V_{ \lambda} \simeq A_i\, V_{ \lambda}\,, \quad i=1,2
\eqno(6.15)  $$
$$
   A_i =
{\Gamma(\langle\lambda+\rho,\delta-\alpha^i-\alpha^0\rangle)
             \over
              \Gamma(\langle\lambda+\rho,-\alpha^0\rangle)}\,.
$$
The first statement in this chain follows, using the Kac-Moody
and ghost algebra commutation relations, from the explicit form
of the corresponding MFF vectors, i.e., (cf. (2.11) or (A.3))
$$
  V_{w_{\beta_1\cdot \zl_1}} = \fb_0^{M^2} \fb_2^{m_1}
  \fb_0^{m_1-M^2} \, V_{\zl_1}\,,
  \quad  \quad V_{w_{\beta_2\cdot \zl_2}} =\fb_0^{M^1}
\fb_1^{m_2}
  \fb_0^{m_2-M^1} \, V_{\zl_2}\,,
$$
and the quantum constraint (6.11), while the second -- from the
quantum constraints (6.12) (see Appendix C for details).  Quite
similarly one shows that the Kac-Moody singular vectors occurring
for $M^3=  -m_3+{2\over \nu}\,, $  $m_3\in \dN\,,$ and related to
the root $\zb= 2 \delta - \alpha^3= w_0(\alpha^3)\,,$
($w_{\beta}= w_{0 1 2 1 0} \in \overline{W}^{(\rho)}\,$)
reproduce trivially the highest weight states in the
corresponding ${\cal W}_3$ Verma modules.

The relations (6.15) provide the basic tool which has to be used
along with the elementary singular vectors (6.8) to construct
the general singular vectors of the ${\cal W}_3\,$  modules.
Indeed recalling that each of the subfactors in the general MFF
monomial represents formally a singular vector we can start as in
the previous section to replace  from the left the powers of the
simple roots vectors $\fb_i\,, i=1,2$ with the corresponding
operator (6.8), while the triples corresponding to the
reflections $w_{010}$ and $w_{020}$ can be simply ``deleted'',
more precisely substituted by the numbers $A_i$ according to
(6.15), reflecting  the implementation of the constraints  $\hat
e^{\alpha^i} \simeq 1$, $i=1,2$.

Let us first illustrate this on the simple example  corresponding
to the root $\beta=\zb_{1, +, 1} = \zd +\za^1$, $w_\beta=w_1 w_0
w_2 w_0 w_1$.  According to the  general formula (A.1) in
Appendix A the Kac-Moody singular vector is given explicitly by
$$\eqalign{
  V_{w_{\zb}\cdot \zl}
  &=  (f^1_0)^{m_1+\kk} \,  (e^3_{-1})^{m_1-\kk+M^2} \,
(f^2_0)^{m_1} \,
     (e^3_{-1})^{\kk-M^2} \,  (f^1_0)^{m_1-\kk} \, \V
\cr\cr
        & = (f^1_0)^{\la w_{0201}\cdot \zl +\zr , \za^1 \ra}
          \,  V_{w_{0201}\cdot \zl} \,,
\cr}\eqno(6.16a)$$
where $m_1\in \dN$.

First we apply (6.6)  to
 the left most power of $f^1_0\,$ and then, accounting for
 (6.15) we  get rid of the central triple of Kac-Moody
generators replacing it by identity
 modulo cohomologically
trivial terms,
(to be more precise, by identity times the numerical
constants $A_i$ -- cohomological equalities up
to numerical coefficients we will denote by $\sim$)
i.e.,
$$
 V_{w_{0201}\cdot \zl}=
 (e^3_{-1})^{m_1-\kk+M^2} \, (f^2_0)^{m_1} \,
     (e^3_{-1})^{\kk-M^2}  \, V_{w_{1}\cdot \zl} \sim
          V_{w_{1}\cdot \zl} \,.
$$

Using once again (6.8), the properties of $Q$ and the
BRST invariance of the $\W_3$ generators (3.7)
(namely $Q$ annihilates Kac-Moody singular vectors
and commutes with ${\cal O}$)
 one obtains
$$
 V_{w_{10201}\cdot \zl}
 \sim
 \cO^{(1)}_{w_{0201} \cdot \zl} \,
  \cO^{(1)}_{ \zl} \, \V \,.
\eqno(6.16b)$$
One has  $h^{(2)}_{w_{10201}
\cdot \zl^{}}=
h^{(2)}_{\zl^{}}+2m_1\,,$ i.e., using (6.1,2) one finds that
the vector (6.16b) provides a singular vector at $L_0\,$ level $2
m_1$
of the $\W_3$ Verma module over $\V$ identified with $ |
h^{(2)}_{\zl^{}}\,, h^{(3)}_{\zl^{}}\ra\,.$

In getting (6.16b) we have worked formally assuming that a proper
analytic continuation of the basic operators (6.8) exists.
One can expand ${\cal O}^{(i)}_\lambda$ as a series in
decreasing powers of $W_{-1}$
(though at first it may seem more natural to expand in  powers
of $L_{-1}$ this is not possible due to the term
$L_{-p}\,{\cal R}^{(i)}\,C_{ij}\,h^j_0$ in (4.5))
$$
   {\cal O}^{(i)}_\lambda =
   {\Gamma(\langle\lambda+\rho,-\alpha^0-\alpha^i \rangle)
    \over\Gamma(\langle\lambda+\rho,-\alpha^0\rangle)}\,
  \sum_{k=0}\, W_{-1}^{M^i-k}\, \sum_{}\,
  c^{(i)}_{k;m_1,\dots,m_s; n_1\dots n_t}
    \, W_{-m_1}\dots W_{-m_s}\,L_{-n_1} \dots L_{-n_t}\,,
\eqno(6.17)$$
where the second sum is over positive sets of integers
 $\{2\le m_1\le\dots\le m_s\}$
and $\{n_1\le\dots\le n_t\}$ such that
$\sum_i m_i + \sum_j n_j = k$.
The coefficients are polynomials in the parameters characterising
the weight, i.e., $M^1$, $M^2$, and ${1\over\nu}$,
so (dropping the overall constant) one can continue analytically
(6.17) to an infinite series in the spirit of Kent [18].
The first coefficient is (up to a sign) $c_0^{(i)}=1$
(absorbing here for simplicity the normalisation $a_{\nu}$ in the
current $W(z)$).
Because of the similarity
with the case in [18] we shall not carry out the analog of the
full
analysis performed there. Instead in  Appendix D  we show
how the universal  coefficients
in this expansion can be computed, giving them
explicitly  up to $k=2$, which is enough to write down the  vector
in example (6.16b)  with $m_1=1$ in a standard, integer powers
form.

Following the same  steps, as in the example above, one obtains
 in general starting from (A.1), (A.3), respectively,
   the following expressions for the $\W_3$ singular vectors
 corresponding to the roots (2.12a,b)
$$\eqalign{
V_{w_{\zb}\cdot \zl} \simeq   \cS_{w_{\beta}\cdot\zl}^{(1)}\,\V
  & \equiv  N_1 \,
    \cO^{(1)}_{(w_{0201})^{m_1'-1}\cdot\zl} \dots
    \cO^{(1)}_{w_{0201}\cdot\zl}\, \cO^{(1)}_{\zl}\,\V \,,
\cr
  {\rm if\ }  M^1
  & = m_1 -  {m_1'-1 \over \nu }\,, \, \, m_1, m_1'\in \dN
  \,,\quad   \beta=(m_1'-1)\,\delta + \alpha^1\,,\cr}
  \eqno(6.18)$$
$$\eqalign{
V_{w_{\zb}\cdot \zl} \simeq
  \cS_{w_{\beta}\cdot\zl}^{(2)}\,\V
    & \equiv  N_2
    \cO^{(2)}_{(w_{0102})^{m_2'-1}\cdot\zl} \dots
    \cO^{(2)}_{w_{0102}\cdot\zl}\, \cO^{(2)}_{\zl}\,\V\,,
\cr
{\rm if\ }  M^2
& = m_2 -  {m_2'-1 \over \nu }\,, \, \, m_2, m_2'\in
\dN \,,\quad    \beta=(m_2'-1)\,\delta + \alpha^2\,,
\cr}\eqno(6.19)$$
where $  N_i$ are numerical constants (see Appendix C).
Similarly in the case (2.12c) one obtains
$$\eqalign{
V_{w_{\zb}\cdot \zl}  \simeq
  \cS_{w_{\beta}\cdot\zl}^{(3)}\,\V
  &\equiv   N_3 \,
  {\cal O}^{(1)}_{ w_{21(0121)^{n}}\cdot\lambda}
  {\cal O}^{(2)}_{ w_{1(0121)^{n}} \cdot\lambda}
  {\cal O}^{(2)}_{ w_{1(0121)^{n-1}} \cdot\lambda}\dots
  {\cal O}^{(2)}_{ w_{1(0121)} \cdot\lambda}
  {\cal O}^{(2)}_{ w_1\cdot\lambda}
  {\cal O}^{(1)}_{ \lambda}\, V_\lambda\,,
 \cr
{\rm if\ }  M^3
& = m_3 -  {n\over \nu }\,, \, n=m_3'-2\,,   \,m_3'-1\,, m_3 \in
\dN
 \,,\quad    \beta= (m_3'-2) \,\delta + \alpha^3\,.
\cr}
\eqno(6.20)$$

According to (6.1-3)  these states (6.18-20) are annihilated by
the positive modes of the $\cW_3$ algebra currents, while their
$L_0$  level  is $ m_1\, m_1'\,,$  $ m_2\, m_2'\,,$ and $ m_3\,
m_3'\,,$ respectively, since $h_{w_{\zb_{n,+,i}}\cdot\zl}^{(2)} =
h_{\zl}^{(2)} +  \la \zl +\zr , \zb_{n,+,i} \ra \, (n+ \la \zr ,
\za^i \ra)\,.$ Formally also any subfactor of these expressions
obtained by deleting from the left the first, the first two,
etc., operators $\cO$,  is a singular vector.

 Since the representations (2.12) for the affine Weyl reflections
are not unique, there are various equivalent ways of representing
the Kac-Moody and hence the ${\cal W}_3$  singular vectors. For
example, the vector in (6.20) for $n $ - even can be also
represented starting from (A.2) as

$$
V_{w_{\zb}\cdot \zl}  \simeq
  \cS_{w_{\beta}\cdot\zl}^{(3)}\,\V
  \equiv   N_3'' \,
  \cO^{(3)}_{ w_{(0121)^{2l}}\cdot\zl}\dots
  \cO^{(3)}_{ w_{(0121)^{2}} \cdot\zl}
  \cO^{(3)}_{ \zl} \, \V\,,
\eqno(6.21a)
$$
where
$$\cO^{(3)}_{ w_{(0121)^{2t}} \cdot\zl}=
\cO^{(1)}_{w_{21 (0121)^{2t}} \cdot\zl}
\cO^{(2)}_{w_{1(0121)^{2t}}\cdot\zl}
\cO^{(1)}_{w_{(0121)^{2t}} \cdot\zl}\,.
\eqno(6.21b)
$$

Furthermore exploiting the symmetry (6.13) the same vectors
(6.18-20) can be recovered starting from the MFF vectors
corresponding to the Weyl reflections $w_{\zb}\,$, with $\zb\,$
as in (2.12d),  (2.12e), or (2.12f), respectively, i.e.,  $\zb =
(m_1'+1)\zd -  \za^1\,,$  $\zb = (m_2'+1)\zd -
\za^2\,,$ or   $\zb = (m_3'+2)\zd -
\za^3\,,$ in the $\hs$ modules  $\zO_{w_{020}\cdot \zl}\,,$
$\zO_{w_{010}\cdot \zl}\,,$ or,  $\zO_{w_{01210}\cdot \zl}\,,$
respectively, since (cf. also (6.14b),  (6.14a), or, (6.14e)
respectively)
 $$\la w_{020}\cdot\zl +\zr ,\,\zb_{n+2, -, 1}
 \ra
 =m_1=
\la \zl +\zr , \,
\zb_{n, +, 1} \ra\,,
$$
$$\la w_{010}\cdot\zl +\zr ,\,\zb_{n+2, -, 2}
 \ra
 =m_2=
\la \zl +\zr , \,
\zb_{n, +, 2} \ra\,,
$$
$$\la w_{01210}\cdot\zl +\zr ,\,\zb_{n+4, -, 3} \ra
 =m_3=
\la \zl +\zr , \,
\zb_{n, +, 3} \ra\,.
$$

 The resulting expressions for the ${\cal W}_3\,$ singular
vectors are identical up to constants to (6.18),  (6.19), or,
(6.20), respectively, i.e.,

$$  \cS_{w_{\beta_{n,+,1}}\cdot\zl}^{(1)}\,\V \sim
  \cS_{w_{\beta_{n+2,-,1}} w_{020}\cdot \zl}^{(1)}\,
  V_{w_{020}\cdot \zl}\,,\eqno(6.22a)
$$
$$  \cS_{w_{\beta_{n,+,2}}\cdot\zl}^{(2)}\,\V \sim
  \cS_{w_{\beta_{n+2,-,2}} w_{010}\cdot \zl}^{(2)}\,
  V_{w_{010}\cdot \zl}\,,\eqno(6.22b)
$$
$$  \cS_{w_{\beta_{n,+,3}}\cdot\zl}^{(3)}\,\V \sim
  \cS_{w_{\beta_{n+4,-,3}} w_{01210}\cdot \zl}^{(3)}\,
  V_{w_{01210}\cdot \zl}\,.\eqno(6.22c)
$$

Apart from the case $\zb =3\zd -\za^3\,,{3\over \zn}-M^3=m_3\in
\dN \,$, we have
enumerated all $\cW_3\,$ singular
vectors resulting from Kac-Moody ones. However the vectors
(6.18-20) are not all independent. Indeed
the symmetry (6.13) implies furthermore that
for $M^1=m-{n \over \nu}\,,$ $m,n+1 \in \dN$,
   the three types of vectors (6.18), (6.19), (6.20),
provide different realisations for
one and the same
(up to an overall constant) singular vector in the $\W_3$ module,
after identifying the Kac-Moody h.w. states with the $\W_3$ one,
more precisely, (cf. (6.14d) and (6.14a) and note that the
symmetry group intertwines the 6 types of real positive roots in
 (2.12))
$$  \cS_{w_{\beta_{n,+,1}}\cdot\zl}^{(1)}\,\V \sim
  \cS_{w_{\beta_{n,+,2}} w_{0210}\cdot \zl}^{(2)}\,
  V_{w_{0210}\cdot \zl} \sim
  \cS_{w_{\beta_{n-1,+,3}} w_{010}\cdot \zl}^{(3)}\,
   V_{w_{010}\cdot \zl}\,,\eqno(6.23)$$
(For $n=0$ only the first relation in (6.23) holds and the
two vectors  reduce to  the elementary singular vectors (6.8)).
Similar relations hold for  $M^2=m-{n \over \nu}\,,$ if $m,n \in
\dN$,
starting with the vector in (6.19), and for  $M^3=m-{n \over
\nu}\,,$ if $m,n+1 \in \dN$, starting with the vector in (6.20).
Finally for $\beta =3 \delta - \alpha^3$ and ${3\over
\nu}- M^3= m_3\in \dN $  (the only case not covered by  the
above) the singular vectors are simply given by $  {\cal
O}^{(1)}_{ w_{0210}\cdot\lambda}\,V_{ w_{0210}\cdot\lambda} $.

Combining (6.22) and (6.23)
  for $\zl$ satisfying, say the
condition in (6.18) ($n\in \dN$),  we can write

$$  \cS_{w_{\beta}\cdot\zl}\,\V \sim
  \cS_{w_{\beta'} \cdot \zl'}\,
  V_{ \zl'}\,, \quad \zl'= w\cdot \zl\,,\ \  \beta' = w(\beta)\,,
\quad w \in \overline{W}^{(\rho)}\,,
\eqno(6.24)$$
where we have used the notation $\cS_{w_{\beta_{n,\pm,
i}}\cdot\zl}:=\cS_{w_{\beta_{n,\pm, i}}\cdot\zl}^{(i)}\,. $ This
implies  that (at least for generic weights, see below) the basic
singular vectors of the type in  (6.18) are enough to reproduce
all $\W_3$ singular vectors.  Let us finally remark that the
relations (6.23) lead to relations between the basic operators
${\cal O}^{(i)}_\lambda$ for $i=1, 2$.  The simplest one
corresponds to the first relation in  (6.23), i.e.,
$$
  {\cal O}^{(2)}_{w_{0210}\cdot\lambda} =
  { \Gamma(\langle\lambda+\rho,2\delta-\alpha^3\rangle) \,
    \Gamma(\langle\lambda+\rho,\alpha^3-\delta\rangle)
   \over
    \Gamma(\langle\lambda+\rho,2\delta-\alpha^2\rangle) \,
    \Gamma(\langle\lambda+\rho,\alpha^2-\delta\rangle) }\,
  {\cal O}^{(1)}_\lambda
$$
and comparing the coefficients in the expansion (6.17)
we check that indeed this relation is true up to the
calculated order (see Appendix D). \footnote{$^2$}{It is not so
straightforward in general to show
explicitly this coincidence, thus the statement relies on the
assumption that at a given weight (6.3)  there is at most one
singular vector in a $\cW_3$ Verma module.}

Once we insert the explicit form (6.17) of the operators ${\cal
O}^{(i)}_\zl $ into (6.18-20), we get the overall constants
${\cal N}_i\,,$
$\,i=1,2,3$, given by the $N_i\,,$
$\,i=1,2,3$ of Eqs. (C.14) and (C.15),  times the
appropriate products of the factors appearing in the r.h.s. of
(6.17).  The result is (up to signs):
$$
  {\cal N}_{i} =
  {\Gamma(\langle\lambda+\rho,\alpha^3 -\delta\rangle -m_{i})
  \over\Gamma(\langle\lambda+\rho,\alpha^3-\delta\rangle
)}\,,\qquad {\rm for} \quad
  M^{i}=m_{i}-{n_{i}\over \zn } ,\quad
  n_{i}\equiv m'_{i}-1\,, \, \,  i=1,2\,,
\eqno(6.25a)
$$

$$
  {\cal N}_3 =
  {\Gamma(\langle\lambda+\rho,\alpha^3 -\delta\rangle -2m_3)
  \over\Gamma(\langle\lambda+\rho,\alpha^3-\delta\rangle
  )}\,,\qquad
  {\rm for} \quad  M^3=m_3-{n_3\over \zn }\,, \quad n_3\equiv
  m'_3-2\,.
\eqno(6.25b)
$$
 In other words ${\cal N}_j$ is the overall constant in
 (6.18-20) when the coefficient of the highest power of
 $W_{-1}$ is normalized to 1, e.g., in (6.18) we have
 ${\cal S}^{(1)}_{w_\beta\cdot\lambda} \, V_\lambda =
 {\cal N}_1 \left( W^{m_1m_1'}_{-1} + \dots\right)$.

Similarly for ${\cal N}'_j,\, j=1,2,3$ -- the overall constants
of the singular vectors corresponding to the roots $n_j \zd
-\za_j\,$  (to be used in the r.h. sides of
(6.22a-c) properly transforming the weights),  we have the same
expressions as in (6.25) with $m_j,\,n_j$ replaced by
$-m_j,\,-n_j$. For $M^i=-m_i+ \kk\,, i=1,2\,,$    $\, {\cal
N}'_i$
reproduce the constants $A_i$ in the BRST relations (6.15).

\bigskip\noindent{\bf 6.3. The Drinfeld-Sokolov reduction and the
${\cal W}_3$ pseudo - Verma modules.}\medskip

The description of the singular vectors  in the $\W_3$  Verma
modules which we obtained exploiting the quantum hamiltonian
reduction  is qualitatively in agreement with the formulae for
the corresponding Kac determinant [12], [30], [31]. In particular
the $\W_3$  singular vectors for $\lambda \,$ such that $M^3=m_3
+\kk\,,$ $\,m_3 \in \dN\,,$ -- a condition which does not
originate from a KK one -- are recovered exploiting the symmetry
(6.12) from the vectors
$$
\cS_{w_1\, w_{010}\cdot\zl}^{(1)}\,   V_{w_{010}\cdot \zl}\,
\sim
\cS_{w_2 \, w_{020}\cdot\zl}^{(2)}\,   V_{w_{020}\cdot \zl}\,
$$
(cf. (6.23)).

  What remains to be understood however, is the exact meaning of
the expressions for the vectors described.  Indeed, up to now we
have worked formally and we have neglected the possible
singularities of the numerical constants (6.25) in the explicit
expressions for the singular vectors.
The  rigourous treatment of these singularities
 requires
 further investigation -- here we will only sketch the
 implications in some simple examples.

Let us start with the simplest case, when there is an overall
constant $1/\epsilon \rightarrow \infty$ in the basic singular
vector (6.8) (cf. (6.17)). Since the l.h.s. of (6.8) is finite,
this singularity  implies that the $\W_3$ singular vector
vanishes identically when we express the $\W_3$  generators back
in terms of the Kac-Moody ones --  and simultaneously identify
the vacuum state in the $\W_3$ module with a Kac-Moody one. In
particular a relative infinite constant might appear in the
relations (6.22), (6.23), so that some of the representatives of
a $\W_3$
vector reproduce it directly, while some provide it only  after
taking the residue of the analytically continued expression.
E.g., consider the example $M^1=m \in \dN, M^2= \kk$,  i.e.,
$M^3= m+\kk$, and assume that  $1/ \nu$  is not an integer in the
interval $[1,m]\,$. According to (6.17) the overall constant in
the first vector in (6.23) is infinite while in the second it is
finite.  Thus this latter vector has to be used to represent the
$\W_3$ singular vector.

The case $ M^1=m=m' + \kk\,,$ $ M^2= \kk\,,$  where $ 1/\nu$ is
an integer in the interval $[1,m]\,,$ in this example is somewhat
more exceptional, since neither of the first two vectors in
(6.23) reproduces directly a well defined expression for the
$\W_3$ singular vector.
Consider the simplest of these
examples -- $(M_1,M_2)= (\kk, \kk)\,. $
  For  the simple roots
$\za^i\,, i=1,2$ we have two different  singular
vectors in the given KM module, which
have the same
eigenvalues $\{h^{(2)}_{\zl} + m,0\} \,$ of the zero modes
$L_0\,, W_0$.

Let us for simplicity illustrate this degenerate case by the
simplest value $m=1$, i.e., by  the example in  (4.1) for
$\kk=1$.   In both linear combinations in (4.1) the  overall
constant $ M^3 -1 - \kk $ in the l.h.s. vanishes along with the
constants in front of $L_{-1}$, while the vector $W_{-1}\V $ is
identically zero (cf. (3.7b)).  Thus we can extract at most one
finite nontrivial linear combination of the two vectors, namely
$(f_0^1+ f_0^2)\V = L_{-1}\V\,,$ while the state $(f_0^1-
f_0^2)\V$, although belonging to the kernel of $Q$, cannot be
expressed in terms of $\W_3$ generators.  The only state selected
$L_{-1}\,\V\,,$  gives rise to a  $\W_3$ (factor) module of
weight $\{  h^{(2)}_{\zl'}\,, h^{(3)}_{ \zl'}\,\}\,,$ where
accounting for the symmetry (6.13,14) $\zl'$ can be taken to be
$\overline{\zl'}+\overline{\zr} = (2  ,2 )\,.$ Furthermore the
composite singular vectors in (6.9) (there are three here since
both $M^1$ and $M^2$ are integers -- $V_{w_{12}\cdot\lambda}$,
$V_{w_{21}\cdot\lambda}$ and $V_{w_{121}\cdot\lambda}$) will be
affected -- namely they have to be built using the finite linear
combination. (Note that they do not involve further singularities
and in particular the weights of the first two vectors
$V_{w_{12}\cdot\lambda}$ and $V_{w_{21}\cdot\lambda}$ lie on
different orbits of the symmetry group.)

There is another alternative of  extracting finite expressions
from the two linear combinations in the r.h.s. of (4.1).  Indeed,
no more assuming the original identification (3.7b) of the $\W_3$
generators with elements in the Kac-Moody enveloping algebra one
can continue analytically in the parameter $\kk =1 + \ze$. Then
by taking proper limits    one can  get from the initial singular
linear combinations a pair of states,

$$\lim_{\ze \rightarrow 0} \ze \, \cS_{\ze}^{(1)} =  W_{-1}\,,
$$
$$\lim_{\ze \rightarrow 0}
 \p_{\ze}\, ( \ze \, \cS_{\ze}^{(1)})
  ={1 \over 2}
L_{-1} \,,\eqno(6.26)
$$
where
$$\cS_{\ze}^{(i)} = {1 \over \ze}(\zd_{i1}-\zd_{i2}) W_{-1}
 + {1 \over 2} L_{-1}\,, \quad \kk=1+\ze\,,
\, \, M^1=M^2=1\,,
$$
neglecting the terms of order $\ze$ or higher.  (There is an
arbitrariness in the second state, namely one can add $b\,
W_{-1}$ with an arbitrary constant $b$;  we have dropped such a
finite term in $\cS_{\ze}^{(1)}$.) The positive mode $\W_3$
generators annihilate both states of the pair, while (6.2)
implies that $W_0\, \cS_{\ze}^{(i)}\,\V =  \ze\,
(\zd_{i1}-\zd_{i2})
\cS_{\ze}^{(i)}\,\V\,,$ up to higher orders in $\ze$,  and hence
 $W_{0}$ acts inhomogeneously on the pair  $\{W_{-1}\,
\V , L_{-1}\,\V\}\,, $
i.e., the would be two singular vectors combine and
provide an indecomposable representation of the subalgebra
$\{L_0, W_0\}\,.$ Thus the pair  $\{W_{-1}\,
\V , L_{-1}\,\V\}\,, $  provides the vacuum state of a
pseudo- Verma module of $\W_3$ (see [12], [32]). Imposing the
invariant condition
$W_{-1}\equiv 0$ one recovers the factormodule built on the state
$ L_{-1}\,\V\,. $ (The first state in (6.26) can be alternatively
recovered directly by the representative in the r.h.s. of
(6.22a), since the corresponding overall constant ${\cal N}'_1$,
determined after (6.25b), is finite and nonzero.)

One can expect that what happens in this example is rather
general.  There may be more than one nontrivial  singular vectors
in a given Kac-Moody module, such that their weights $\lambda\,$
lie on one and the same orbit of the invariance group, and such
that both reduced expressions have overall singularities. More
generally,  identifying the KM highest weight states lying on an
orbit of the stability group,
the formal doubling of the singular vectors in the resulting
$\W_3$ Verma module may originate from different KM Verma modules.
In the  series $(m,\kk)\,$ discussed above this can happen for
$m=m'+\kk > \kk\,$
-- e.g., for n=0 the first two vectors in (6.23) originating from
the KM modules labelled by $(m'+\kk , \kk)\,$ and $
(-m'+\kk , m'+ \kk)\,,$ (and having a $\W_3$ weight, which can be
parametrised by $(M_1,M_2)=(m'+{2\over \nu} , {2\over \nu})\,$)
both have
singularities.
Similar procedures as in the example above could be used
in general to provide the  explicit expressions of the $\W_3$
pseudo - singular vectors.

Note that for $ \kk \in \dN \,$ and $(M^1, \kk)\,, \, M^1,  \in
\dN, M^1 > \kk \,$ (or $(\kk, M^2)\,, \, M^2,  \in \dN, M^2 > \kk
\,$) and furthermore for the more general series described by
$(M^1, M^2)\,, \, M^i, \kk \in \dN, M^i > \kk \,$, the orbit of
the stability group effectively splits  into two unrelated sets.
 (The weights $(\kk,\kk)\,$ are invariant with
respect to the symmetry (6.14).)
Indeed, examining the constants in the BRST relations (6.15) with
the  relevant KK conditions taken into account, one sees that
some of them vanish, so that the BRST equivalence of  the
corresponding pairs fails.  The above mentioned identification,
producing the   highest weight state of a ${\cal W}_3$ Verma
module, refers to   states on  one of these  ``half-orbits''
-- the one containing the dominant weight $(M^1,M^2)\,.$ On the
other hand the weights of the two singular vectors with overall
singularities, which
eventually give rise to an indecomposable pair, lie on
a``half-orbit'' of the second type.  The latter contains in
particular $({2 \over \nu} - \tilde{M}^2,   {2 \over \nu} -
\tilde{M}^1)$ (cf. (6.14e)), where $(\tilde{M}^1,\tilde{M}^2)\, $
is a weight of one of the above types.

In the recent paper [32] the weights and the embedding pattern
for some examples of such generalised Verma modules of $\W_3$
(for $1/\nu =1$) were analysed by use of computer computations.
The weights listed are in agreement with the expected Kac-Moody
origin of the pseudo- singular vectors.  Moreover  comparing the
results  in [32] with the predictions provided by the quantum DS
approach
leads us to the conjecture that   the cases   described above
provide  the natural extension of the series of $\W_3$ Verma
modules with  pseudo-singular vectors in [32].
At this stage however it is not clear whether the above series
exhaust all such modules, since in particular the class of
weights for which the first two vectors in (6.23) both have
singularities is larger.

%%%%%%%%%%%%%%%%%%%
%
%     section 7
%
%%%%%%%%%%%%%%%%%%%

\bigskip\noindent{\bf 7. DISCUSSION.}
\medskip

For any Kac-Moody algebra $g$ and any
embedding of $sl(2)$ into the horizontal subalgebra
$\overline g$ (characterized by the defining vector $\eta$ of
the embedding) one has a quantum Drinfeld-Sokolov
reduction leading to a ${\cal W}(\overline g,\eta)$ algebra [11].
Here, on the example of $g=A^{(1)}_2$, we have described a
formalism which is a step towards an understanding
of the Verma module resolutions of the irreducible highest weight
representations of ${\cal W}(\overline g,\eta)$.  Although the
method lacks the  rigour of a complete  cohomological analysis it
is more constructive in that it  provides explicit
expressions for Verma modules singular vectors.

There are essentially two basic ingredients in the method
presented here.
Both can be traced to their classical counterparts.

The first  important ingredient used is the
stability group $\overline W^{(\eta)} $ of the eigenvalues of the
zero modes subalgebras, i.e., of the weights of the ${\cal W}$
algebra.  This is a finite subgroup of the affine Weyl group,
specific for the given $sl(2)$ embedding.  The ${\cal W}$ algebra
singular vectors are governed by $W$ modulo $ \overline
W^{(\eta)}$, much in the same way as the affine Weyl group
classifies the Kac-Moody singular vectors and gives their
embedding pattern.  This  is demonstrated rather  constructively
in our procedure. Indeed the initial (powers of) root vectors
$\{f_0^{\alpha^i}\,, i=1,2,.., r= {\rm
rank} \ \overline g\,; $ $\, e_{-1}^{\theta}\}$,
($\theta$- the highest root) in the general Malikov-Feigin-Fuks
monomial,
corresponding to the generators of the affine Weyl group
$\{w_{\alpha^i}\,, i=1,2,..,r\,;$ $\, w_{\delta - \theta}\}\,$
are  regrouped to arrive at a subset of $\{f_0^{\alpha^i} \,,
i=1,2,..,r\}\,$ and  $\{ e_{-1}^{\alpha}\,, \alpha \in
I^{(\eta)} \}$ ,  the latter corresponding to the generators
$\{w_{\delta -\alpha}\,, \, \alpha \in   I^{(\eta)}\}$ of the
symmetry group.  Here $  I^{(\eta)} $ is the set of roots of
$\bar g$ for which $ e^{\alpha}=1$   (classically).  This
suggests the form of the stability group in general, as it is
confirmed, e.g., by the case of the principal embedding for $\bar
g =sl(n)\,$ (see also [33] for the case of finite ${\cal W}$
algebras).

Secondly, we have exploited a quantum gauge
transformation,   playing a role
quite analogous to the classical Drinfeld-Sokolov gauge fixing
transformations.
Because we have used the Malikov-Feigin-Fuks
description of Kac-Moody singular vectors it was sufficient to
have the projections ${\cal R}^{(i)}$
of the gauge transformations along the
directions of the simple roots $\alpha^i$, $i=1,\dots, r\,.$
The classical analogs might give a clue for finding these quantum
operators  in general.
What seems however
still not quite satisfactory is the rather technically involved
derivation of the intertwining relation in the principal
embedding case
(see Appendix B), which we are not able at present to simplify
considerably.
This reflects the intrinsic asymmetry of our approach -- as
compared with the fusion method initiated in  [24],
in treating the pairs of labels $(m_i, m'_i)$ in the Kac-Kazhdan
conditions.
A more ambitious problem would be to find the general quantum
gauge transformation, a multi series involving also the negative
modes of the raising currents $e^{\alpha_i}\,, i=1,2$,  which we
expect to reproduce the ${\cal W}$ algebra singular vectors
directly.
This is an open problem even for the case
$\overline g=sl(2)$. A related problem is the adaptation in the
standard BRST framework and the
generalisation of the algorithm in [19] for constructing  the
general singular vectors, which starts from a simplified
expression, BRST equivalent to the initial KM singular vector,
and in particular makes unnecessary the use of Kent --
type expressions.

While in the example of $\W_3$ there are complications due to the
appearance of indecomposable pairs of singular vectors at some
values of the parameters $M^1$, $M^2$, and $\nu$, in  the ``less
constrained'' case of the Polyakov-Bershadsky algebra there is a
striking similarity of the expressions for the
singular vectors with those for the initial affine algebra.
In this case the stabilising
group is $\{1,w_0=w_{\delta- \theta}\}$ and the embedding pattern
and hence the
characters are expected to follow in a straightforward manner
from the
corresponding information about the Kac-Moody algebra.

There are other possible applications of the results found here.
In the case of $\widehat sl(2)$ we have a full understanding also
of the reduction of the general correlation functions and the
reduction of the Knizhnik-Zamolodchikov equation (combined with
the algebraic equations resulting from the Malikov-Feigin-Fuks
vectors) to the null-decoupling equations of
Belavin-Polyakov-Zamolodchikov [14], [19]. It would be
interesting to  generalise these results for the ${\cal W}\, $
algebras related to  $\widehat sl(3)$. In particular it is
expected that  differential equations originating from the
Knizhnik-Zamolodchikov equations can be found accounting also for
the ${\cal W}$- algebras singular vectors.  The simplest
differential equations corresponding to the fundamental
representations of $sl(3)$ were considered in [34], [35].

%%%%%%%%%%%%%%%%
%
%     appendix A
%
%%%%%%%%%%%%%%%%

\bigskip\noindent{\bf Appendix A. Malikov-Feigin-Fuks vectors of
$A^{(1)}_2$.}
\medskip

In this appendix we describe in more details  the
Malikov-Feigin-Fuks vectors for $A^{(1)}_2$.
As usual we parametrize a heighest weight $\lambda$ by
$M^i=\langle\lambda+\rho,\alpha^i\rangle$, $i=1,2$.
For the six types of positive roots $\beta$  of $A^{(1)}_2$
using the decomposition (2.12) of  $w_\beta$ into simple
reflections
one gets  from (2.11) the following  expressions.

For $M^1=m-{n-1\over\nu}$, $m,n\in\dN$,  $M^2$ arbitrary
and $ \beta = (n-1)\,\delta + \alpha^1$  one has
$w_\beta=w_1(w_{0201})^{n-1}$
and the Malikov-Feigin-Fuks vector is
$$
  V_{w_\beta\cdot \lambda} =
  \fb_1^{\,A_{n-1}}\,
  \underbrace{ \fb_0^{\,D_{n-2}}\, \fb_2^{\,C_{n-2}}\,
                       \fb_0^{\,B_{n-2}}\, \fb_1^{\,A_{n-2}}  }
  \,\,\dots\,\,
  \underbrace{ \fb_0^{\,D_0}\,\, \fb_2^{\,C_0}\,\,
                       \fb_0^{\,B_0}\,\, \fb_1^{\,A_0} } \,V_\lambda\,,
\eqno(A.1)$$
where
$$\eqalign{
  A_p  &\equiv
  \langle w_{(0201)^p} \cdot \lambda + \rho, \alpha^1\rangle
  = M^1 + {2p\over\nu}                    \,,
\cr\cr
  B_p  &\equiv
  \langle w_{1(0201)^p} \cdot \lambda + \rho, \alpha^0 \rangle
  = - M^2 + {p+1\over\nu}       \,,
\cr\cr
  C_p  &\equiv
  \langle w_{01(0201)^p} \cdot \lambda + \rho, \alpha^2 \rangle
  = M^1 + {2p+1\over\nu}        \,,
\cr\cr
   D_p  &\equiv
  \langle w_{201(0201)^p} \cdot \lambda + \rho, \alpha^0 \rangle
  = M^1 + M^2 + {p\over\nu}  \,.
\cr}$$

For $M^1$ arbitrary, $M^2=m-{n-1\over\nu}$, $m,n\in\dN$,
and $ \beta = (n-1)\,\delta + \alpha^2$  one has
$ w_\beta = w_2(w_{0102})^{n-1} $ and
the Malikov-Feigin-Fuks vector has the same form as
the one above except that ``1'' and ``2'' are interchanged.

For $ M^1+M^2 = m - {n-1\over\nu}$, $m,n\in\dN$
and $ \beta = (n-1)\,\delta + \alpha^1 + \alpha^2 $
one has $ w_\beta = w_{121} (w_{0121})^{n-1} $ and the
Malikov-Feigin-Fuks vector is
$$
  V_{w_\beta\cdot \lambda} =
  \fb_1^{\,C_{n-1}}\,\fb_2^{\,B_{n-1}}\, \fb_1^{\,A_{n-1}}\,
  \underbrace{ \fb_0^{\,D_{n-2}}\, \fb_1^{\,C_{n-2}}\,
                       \fb_2^{\,B_{n-2}}\, \fb_1^{\,A_{n-2}}  }
  \,\,\dots\,\,
  \underbrace{ \fb_0^{\,D_0}\,\, \fb_1^{\,C_0}\,\,
                       \fb_2^{\,B_0}\,\, \fb_1^{\,A_0} }
\,V_\lambda\,,
\eqno(A.2)$$
where
$$\eqalign{
  A_p  &\equiv
  \langle w_{(0121)^p} \cdot \lambda + \rho, \alpha^1\rangle
  = M^1 + {p\over\nu}                    \,,
\cr\cr
  B_p  &\equiv
  \langle w_{1(0121)^p} \cdot \lambda + \rho, \alpha^2 \rangle
  = M^1 + M^2 + {2p\over\nu}       \,,
\cr\cr
  C_p  &\equiv
  \langle w_{21(0121)^p} \cdot \lambda + \rho, \alpha^1 \rangle
  = M^2 + {p\over\nu}        \,,
\cr\cr
   D_p  &\equiv
  \langle w_{121(0121)^p} \cdot \lambda + \rho, \alpha^0 \rangle
  = M^1 + M^2 + {2p+1\over\nu}  \,.
\cr}$$

For $M^1=-m+{n\over\nu}$, $m,n\in\dN$,  $M^2$ arbitrary
and $ \beta = n\,\delta - \alpha^1$  one has
$w_\beta=w_{020}(w_{1020})^{n-1}$
and the Malikov-Feigin-Fuks vector is
$$
  V_{w_\beta\cdot \lambda} =
  \fb_0^{\,C_{n-1}}\, \fb_2^{\,B_{n-1}}\, \fb_0^{\,A_{n-1}}\,
  \underbrace{ \fb_1^{\,D_{n-2}}\, \fb_0^{\,C_{n-2}}\,
                       \fb_2^{\,B_{n-2}}\, \fb_0^{\,A_{n-2}}  }
  \,\,\dots\,\,
  \underbrace{ \fb_1^{\,D_0}\,\, \fb_0^{\,C_0}\,\,
                       \fb_2^{\,B_0}\,\, \fb_0^{\,A_0} }
\,V_\lambda\,,
\eqno(A.3)$$
where
$$\eqalign{
  A_p  &\equiv
  \langle w_{(1020)^p} \cdot \lambda + \rho, \alpha^0\rangle
  = - M^1 - M^2 + {p+1\over\nu}                    \,,
\cr\cr
  B_p  &\equiv
  \langle w_{0(1020)^p} \cdot \lambda + \rho, \alpha^2 \rangle
  = - M^1 + {2p+1\over\nu}       \,,
\cr\cr
  C_p  &\equiv
  \langle w_{20(1020)^p} \cdot \lambda + \rho, \alpha^0 \rangle
  = M^2 + {p\over\nu}        \,,
\cr\cr
   D_p  &\equiv
  \langle w_{020(1020)^p} \cdot \lambda + \rho, \alpha^1 \rangle
  = - M^1 + {2p+2\over\nu}  \,.
\cr}$$

For $M^1$ arbitrary, $M^2=-m+{n\over\nu}$, $m,n\in\dN$,
and $ \beta = n\,\delta - \alpha^2$  one has
$ w_\beta = w_{010}(w_{2010})^{n-1} $ and
the Malikov-Feigin-Fuks vector has the same form as
the one above except that ``1'' and ``2'' are interchanged.

For $ M^1+M^2 = - m + {n\over\nu}$, $m,n\in\dN$
and $ \beta = n\,\delta - \alpha^1 - \alpha^2 $
one has $ w_\beta = w_0 (w_{1210})^{n-1} $ and the
Malikov-Feigin-Fuks vector is

$$
  V_{w_\beta\cdot \lambda} =
  \fb_0^{\,A_{n-1}}\,
  \underbrace{ \fb_1^{\,D_{n-2}}\, \fb_2^{\,C_{n-2}}\,
                       \fb_1^{\,B_{n-2}}\, \fb_0^{\,A_{n-2}}  }
  \,\,\dots\,\,
  \underbrace{ \fb_1^{\,D_0}\,\, \fb_2^{\,C_0}\,\,
                       \fb_1^{\,B_0}\,\, \fb_0^{\,A_0} }
\,V_\lambda\,,
\eqno(A.4)$$
where
$$\eqalign{
  A_p  &\equiv
  \langle w_{(1210)^p} \cdot \lambda + \rho, \alpha^0\rangle
  = - M^1 - M^2 + {2p+1\over\nu}                    \,,
\cr\cr
  B_p  &\equiv
  \langle w_{0(1210)^p} \cdot \lambda + \rho, \alpha^1 \rangle
  = - M^2 + {p+1\over\nu}       \,,
\cr\cr
  C_p  &\equiv
  \langle w_{10(1210)^p} \cdot \lambda + \rho, \alpha^2 \rangle
  = - M^1 - M^2 + {2p+2\over\nu}        \,,
\cr\cr
   D_p  &\equiv
  \langle w_{210(1210)^p} \cdot \lambda + \rho, \alpha^1 \rangle
  = - M^1 + {p+1\over\nu}  \,.
\cr}$$

Now we describe an algoritm that transfoms a Malikov-Feigin-Fuks
monomial, in which the generators are raised in general to
complex powers and thus a priori is not an element
of the universal enveloping algebra,
into an ordinary polynomial of elements of ${\tt n}_-$ and
thus an element of the univerlsal enveloping algebra. First note
that all the above monomials have the form
$$
  \dots \,\fb_{i_{2}}^{\,\gamma_{2}} \,
\fb_{i_{1}}^{\,\gamma_{1}} \,
  \fb_{i_{0}}^{\,\gamma_{0}} \, \fb_{i_{1}}^{\,\gamma_{-1}} \,
  \fb_{i_{2}}^{\,\gamma_{-2}} \, \dots
\eqno(A.5)$$
where $\fb_{i_0}^{\,\gamma_0}$ is the middle term (a Weyl
reflection $w_\beta$ always decomposes into an odd number
of simple reflections) and moreover $\gamma_0$,
$\gamma_i+\gamma_{-i}$, are positive integers.
Next note that it is straightforward to continue
analitically in $\gamma$ commutation formulas
of the sort
$$
  [\fb_0^{\,\gamma}\,, f^1_{0}]
  = -\gamma \, e^2_{-1}\, \fb_0^{\,\gamma-1}
\eqno(A.6)$$
or more generally
$$
  \fb_0^{\,\gamma}\, (f^1_{0})^n =
  \sum_{k=0}^n (-1)^k {n\choose k} {\gamma!\over(\gamma-k)!}
  (f^1_0)^{n-k} \, (e^2_{-1})^k \, \fb_0^{\,\gamma-k}.
\eqno(A.7)$$
Thus the algoritm is, using the above and their analogs,
{\it start from the middle and move outwards}.
For example take (A.1) with $n-1=2r$ even.
In this case $i_0=1$, $i_1=0$, $i_2=2$, $i_3=0$, $i_4=1$,
etc., and $\gamma_0=A_r=m$,
$\gamma_1=B_r=-M^2+{r+1\over\nu}$,
$\gamma_{-1}=D_{r-1}= m + M^2 -{r+1\over\nu}$, etc.
Thus using (A.7) we transform the three middle terms of
(A.5) into an ordinary polynomial. Using the analogs
of (A.7) we can continue moving outwards.

%%%%%%%%%%%%%%%%%
%
%   appendix B
%
%%%%%%%%%%%%%%%%

\bigskip\noindent{\bf Appendix B.  Derivation of the basic
relation (4.5)}
 \medskip

In this appendix we will derive the key relation (4.5).

On both sides of (4.5) we have a vector $V$ such that
$$
  e^i_n \, V =  f^i_n \, V =  h^i_n \, V = b^i_n \, V =  c^i_n \,
V =
  c^i_0 \, V = 0
  \qquad \forall n\ge 1, i=1,2,3\,,
\eqno(B.1)$$
and thus $L_n \, V = \Lf_n \, V = 0$ for $n\ge 1$ and
$  \he^i_0 \, V = e^i_0 \,V \,$,
$  \hf^i_0 \, V = f^i_0 \,V \, $,
$  \hh^i_0 \, V = h^i_0 \,V \, $.

To be more concrete we will
consider the intertwining property of the gauge transformation
projected on the ``$1^{\rm st}$ direction, i.e., $\R^{(i)}$ with
$i=1$.
For notational simplicity set
$$ \HH(z) = C_{1i}\hh^i(z) \,. \eqno(B.2)$$
and skip the superscript of $\R$, i.e.,
our gauge transformation (4.2) is
$$
  \R(u) = \exp({\scriptstyle\int} \HH_{(-)}(-u)\,du)
  =\sum_{n=0}^{\infty} \R_{-n} \, u^n \,.
\eqno(B.3)$$
Though we should set $u=e^1_0$
 for the moment let us view it as a scalar
parameter.
Taking derivatives in $u$ we immediatelly obtain
$$\eqalign{
  n\,\R_{-n} &= \sum_{k=0}^{n-1} (-1)^{n-k+1} \R_{-k}\,\HH_{-n+k}
\,,
\cr\cr
  n(n-1)\,\R_{-n} &= \sum_{k=0}^{n-2} (-1)^{n-k+1} \R_{-k} \,
   \left( \p\HH -  \HH\diamond\HH \right)_{-n+k} \,,
\cr\cr
  n(n-1)(n-2) \, \R_{-n} &= \sum_{k=0}^{n-3} (-1)^{n-k+1} \R_{-k}
\,
  \left( (\HH^{\diamond 3}) - {3\over 2} (\p\HH^{\diamond 2}) +
\p^2 \HH
  \right)_{-n+k} \,,
\cr}\eqno(B.4)$$
where for short  $\diamond$  indicates the corresponding normal
product in which only negative mode generators are retained.

In fact the modes $ \R_{-n} \,$ of the gauge transformation
 can be defined recursively by $\R_0=1$
and the first of the equalities $(B.4)$. Then the other
equalities are
obtained by induction.

First we will give some motivation for what we will be doing
below.
In the  $sl(2)$ case the derivation of the corresponding key
relation
can be devided roughly into two steps: 1) one commutes the
 powers of the gauge generator $(e^1_0)^n$
 through the zero mode of the lowering operator $f_0$
producing factors $n$ and $n(n-1)$ which by (B.4) are
transformed into first and second powers of
the Heisenberg field and thus one recovers the free field (f{}f)
part the stress energy tensor; 2) commuting the modes $\R_{-n}$
through $f_0$ one obtains lower modes of $f$ which together with
the (f{}f) piece give the reduced energy momentum tensor. In the
$sl(3)$ case we will follow analogous steps. Since in the example
we are considering we want to move along the ``$1^{\rm st}$
simple direction'' it is natural to take a combination of the
reduced generators $W$ and $T$ (2.7) which eliminates $f^2$.
Denote
$$ {\cal X}(z) =
 \left( \kk\,(\HH\,T) +
 {1\over 2\nu}\, \big(\kk-1\big) \,\p T + \aw\,W \right)(z) \,.
\eqno(B.5)$$
{}From (3.6) and (3.7) we have
$$
  {\cal X}(z) - {\cal X}^{\rm f{}f}(z) =
  \left( (\hh^3\,\hf^1) + \big(\kk-1\big)\,\p\hf^1 +
\hf^3\right)(z) \,.
\eqno(B.6)$$
with
$$
  {\cal X}^{\rm (f{}f)}(z) =
 \left( (\HH(\HH^2)) + {3\over 2}\big(\kk-1\big)\,(\p\HH^2) +
  \big(\kk-1\big)^2 \, \p^2\HH \right)(z) \,.
\eqno(B.7)$$
The fact that
${\cal X}^{\rm (f{}f)}(z)$ depends only on the Heisenberg field
$\HH(z)$
is a good sign and we can hope to recover it from derivatives of
$(B.3)$.
Taking $\oint dz \, z\,\ $ of the r.h.s. of $(B.6)$ one obtains
$
  (\hh^3\,\hf^1)_0 + \big(\kk-1\big)\, (\p\hf^1)_0 + (\hf^3)_1
$
which applied to a vector $V$ with properties $(B.1)$  gives
$$
  \big(h^3_0 + 2 -\kk\big) \, \f10  \, V \,.
\eqno(B.8)$$
This will be our starting point, namelly we apply $\R$ to $(B.8)$
and move it to the right.
There is another motivation for starting with $(B.8)$ (i.e., not
replacing $h_0^3$ with the corresponding numerical eigenvalue)
 -- in the $sl(3)$ case the
(f{}f) part of the reduced generators is third order in the
Heisenberg
field so extending the $sl(2)$ case we will need factors $n$,
$n(n-1)$
and $n(n-1)(n-2)$ which could be obtained from the
commutation of $(e^1_0)^n$ through the quadratic
(in the KM generators) expression of (B.8).

Now we start with step 1) of the calculation.
One immediately gets
$$\eqalign{
  [ (e^1_0)^n &, (\hat h^3_0 + 2 - \kk)\,f^1_0 ] =
  - n \, f^1_0 \, (e^1_0)^n
\cr\cr
  &- \left\{ -n(n-1)(n-2) + \big(3H_0-\kk\big) \, n(n-1) -
    \big(\hat h^3_0 + 1 - \kk\big) h^1_0 \, n \right\}
(e^1_0)^{n-1} \,.
\cr}\eqno(B.9)$$
First consider the expression in curly brackets
in the r.h.s. of $(B.9)$. The terms with $n$, $\,n(n-1)$ and
$n(n-1)(n-2)$ effectively give first, second and third derivative
of
$\R$, i.e., formulae $(B.4)$. Moreover when applied on
$V$ with the property $(B.1)$ we have
$$\eqalign{
  (\HH^3) \, V &= \big( (\HH^{\diamond 3}) + 3\, \HH_0
(\HH^{\diamond 2})
                      + 3 \, (\HH_0)^2\,\HH \big) \, V \,,
\cr\cr
  (\p\HH^2) \, V &= \big(  (\p\HH^{\diamond 2}) + 2\,\HH_0\,\p\HH
                      - 2\,\HH_0\,\HH \big) \, V \,,
\cr\cr
  (\HH^2) \, V &= \big(  (\HH^{\diamond 2}) +  2\,\HH_0\,\HH
\big) \, V \,,
\cr}\eqno(B.10)$$
and after some algebra we obtain
$$
  {\cal R}_{-n}\,
  [ (e^1_0)^n, (\hat h^3_0+2-{1\over\nu})\,f^1_0 ] \, V =
  - n \,{\cal R}_{-n}\, f^1_0 \, (e^1_0)^n \, V +
  \sum_k (-1)^{n-k-1}\,{\cal R}_{-k}\,{\cal
Y}_{-n+k}\,(e^1_0)^{n-1} \,, V
\eqno(B.11)$$
where
$$\eqalign{
  {\cal Y}_{-n} &=
  \Big( (\HH^3) - {3\over 2}\,(\p\HH^2) + \p^2\HH \Big)_{-n}
\cr\cr
  &- 3 (\HH_0)^2\,\HH_{-n} - \kk (\HH)_{-n} + \kk\,\p\HH_{-n}
  - \kk\,\HH_0\,\HH_{-n} + \hh^3_0\,(\hh^1_0 + \kk -1)\,\HH_{-n}
\,.
\cr}\eqno(B.12)$$

Until now we have moved only $u=\e10\,$
to the right while the modes
$\R_{-n}$ are still to the left of the Heisenberg field.
To move also the modes we will need
formulae of the type
$$\eqalign{
  \sum_{n=0}^\infty\R_{-n}\, \sum_{p=1} (\HH^2)_{-p} \, u^{n+p-1}
&=
  \sum_{p=1} \left( (\HH^2) + 2\,\cn \,\p\HH\right)_{-p} \,
u^{p-1}
  \,\, \R \,,
\cr\cr
  \sum_{n=0}^\infty\R_{-n}\, \sum_{p=1} (\p\HH^2)_{-p} \,
u^{n+p-1} &=
  \sum_{p=1} \left( (\p\HH^2) - 2\,\cn \,\p\HH
   + \cn \,\p^2 \HH \right)_{-p} \, u^{p-1} \, \R  \,,
\cr\cr
  \sum_{n=0}^\infty\R_{-n}\, \sum_{p=1} (\HH^3)_{-p} \, u^{n+p-1}
&=
  \sum_{p=1} \left( (\HH^3) + 3\,\cn\,(\p\HH^2) + 3\cn\,(\HH^2)
\right.
\cr\cr &\hskip 1cm   \left.
  + {3\over 2} \Big(\cn\Big)^2 \,\p^2\HH\right)_{-p} \, u^{p-1}
\,\R\,,
\cr}\eqno(B.13)$$
which are proved by  induction
using  the recursive definition of $\R$ and
$$
  [\R_{-n},\HH_m] =
  \cn \sum_{k=0}^{n-1} (-1)^{n-k} \, \delta_{k,n-m} \R_{-k} \,.
\eqno(B.14)$$
while the last formula is derived  by induction from
$
  [\HH_m, \HH_n] = \cn \, m \, \delta_{m,-n} \,.
$

Before  proceeding we prepare another formula -- $(B.17)$. First
from the definition of  $\Lf$
$$
  [\Lf_n, \HH_{-m}] = m \HH_{n-m}
  - \Big(\kk-1\Big)\,m(m+1)\,\delta_{n,m}
\eqno(B.15)$$
from where,   again by induction, follows
$$\eqalign{
  [\Lf_m, \R_{-n} ]
  &= \sum_{k=0}^{n-1} (-1)^{n-k+1}\, \R_{-k}\,\HH_{m-n+k}
\cr\cr
  &+ \sum_{k=0}^{n-1} \left( (m-1)\,{C_{12}\over\nu}
   + (m+1) \, \Big(\kk-1\Big)\right)
  (-1)^{n-k} \, \delta_{k,n-m} \, \R_{-k} \,.
\cr}\eqno(B.16)$$
Exploiting the above and the fact that
 positive modes of $\Lf$ annihilate $V$, see $(B.1)$, we obtain
$$\eqalign{
  &-\sum_{n=1} \left( \sum_{k=0}^\infty
  \HH_{-n-k}\,\Lf_k \right) \, u^{n-1}   \R \, V  =
  \R \, \sum_{n=1} \left( \kk\,(\HH^2) +  {1\over 2\nu}
\,(\p\HH^2)
  +  \kk\Big({1\over 2}- \cn\Big) \,\p^2\HH \right.
\cr\cr
  &\hskip 2 cm  + \left. {2\over\nu}\Big(1-\kk\Big)\,\p\HH
  + \hh^3_0\,\Big(\hh^1_0+\kk-1\Big)\,\HH
    -\kk\,\HH_0\,\HH - 3 (\HH_0)^2\,\HH  \right)_{-n}
  \, u^{n-1} \, V \,.
\cr}\eqno(B.17)$$

With the help of $(B.13)$ and $(B.17)$ we can move
the modes $\R_{-n}$ to the right of the
Heisenberg fields in the second term in the r.h.s. of
 $(B.11)$
$$
  \sum_{n=0} {\cal R}_{-n} \,
  \sum_{p=0} {\cal Y}_{-p} \, (-e^1_0)^{p-1}(e^1_0)^n\, V=
  \sum_{p=0} {\cal Z}_{-p} \,{\cal R}\, (-e^1_0)^{p-1} \, V
\eqno(B.18)$$
where
$$\eqalign{
  {\cal Z}_{-n} &=
  \Big( (\HH^3) + {3\over 2}\big(\kk-1\big)\,(\p\HH^2)
  + \big(\kk-1\big)^2\, \p^2\HH \Big)_{-n} \cr\cr
   &+ \kk \Big({1\over 2}\,\p^2\HH + \p\HH \Big)_{-n}
  - \kk\,\sum_{k=0}^\infty \HH_{-n-k}\,\Lf_k \,.\cr}
\eqno(B.19)$$

Let us recapitulate the result till now -- we have
$$
  \R \, (\hh^3_0 + 2 - \kk)\,\f10 \, V =
  \sum_{n=0}^\infty \R_{-n}\, \Big(\hh^3_0 + 2 - \kk-n\Big)
   \,\f10 (\e10)^n \, V
  + \sum_{n=0} {\cal Z}_{-n}\,\R\,(-\e10)^{n-1}\,V \,.
\eqno(B.20)$$

The first term in $(B.19)$ is exactly $\Xf_{-n}$
but note that in the expressions
$ {\cal X}(z) $ and $ {\cal X}^{\rm (f{}f)}(z)$
we have the products $(\HH T)(z)$ and $(\HH\Tf)(z)$ which,
when $\R$ is moved to the right,
should not survive but instead should be expressible
in terms of $T$ and $\Tf$. Thus let us rewrite $(\HH\Tf)(z)$
$$
  (\HH\Lf)_{-t} = \sum_{k=0}^{t-1} \Lf_{-t+k}\,\HH_{-k}
  -  \Big( {1\over 2}\, \p^2\HH + \p\HH \Big)_{-t}
  + \sum_{k=0}^\infty \HH_{-t-k}\,\Lf_k
  + \sum_{k=0}^\infty \Lf_{-t-k}\,\HH_k \,.
\eqno(B.21)$$
With the help of $(B.14)$ we obtain
$$
  \sum_{p=0} \left(\sum_{k=1}^\infty \Lf_{-p-k} \, \HH_k\right)
  \,\R\,  u^{p-1}\, V =
  - \sum_{p=0} \cn \left(\Lf + \p\Lf\right)_{-p} \,\, \R\,
u^{p-1}
  \, V \,
\eqno(B.22)$$
and substituting all into ${\cal Z}$ we find
$$\eqalign{
    \sum_{n=0} {\cal Z}_{-n}\,\R\,(-\e10)^{n-1}\,V
    = \sum_{n=0} \Big( \aw\,\Wf_{-n}
   &- {C_{11}\over \nu^2}\, \Lf_{-n}
   - {1\over 2} \big(1+\kk(2C_{11}-1)\big) (\p\Lf)_{-n}
\cr\cr
   &+  \sum_{k=0}^{n-1} \Lf_{-n+k} \, \HH_{-k} \Big)
   \R \, (-\e10)^{n-1} \, V \,.
\cr}\eqno(B.23)$$
The last term of the above,
using $[(-\e10)^p,\HH_0] = -p\,(-\e10)^{p-1}$, can be rewritten as
$$
   \sum_{n=0} \left( \sum_{k=0}^{n-1} \Lf_{-n+k} \, \HH_{-k}
\right)
   \R \, (-\e10)^{n-1} \, V =
   \sum_{n=0} \Lf_{-n} \, \R \, \HH_0  \, (-\e10)^{n-1} \, V \,.
\eqno(B.24)$$
Recapitulating once again the sum with ${\cal Z}$ in $(B.20)$
exactly recovers the free field
part of the key intertwining formula (4.5) and thus we have
completed step 1) of our calculation.

Step 2) of the calculation basically involves moving the modes
$\R_{-n}$
through $f^1_0$ in the r.h.s. of $(B.20)$ and
for this purpose we need
$$
  \R_{-n} \, \f10 = \sum_{k=0}^n (-1)^{n+k} f^1_{-n+k} \, \R_{-k}
\eqno(B.25)$$
which follows from the Kac-Moody commutation relations and
induction.
Using $(B.24)$, $(B.4a)$ and several resumations for the first
term in the r.h.s. of $(B.20)$ we obtain
$$
\sum_{n=0}^\infty \R_{-n}\, \Big(\hh^3_0 + 2 - \kk-n\Big)
   \,\f10\, u^n \, V =
 \sum_{n=0}^\infty
 \left[  \Big(\hh^3_0 + 1 - \kk\Big) \,
   \hf^1_{-n} - (\p f^1)_{-n}
  + ( \hf^1\diamond\HH )_{-n}  \right] \, \R \,(-u)^n\, V
\eqno(B.26)$$

Equation $(B.24)$ holds also for $T$, i.e.,
$$
  \sum_{n=0} ((L-\Lf)\diamond\HH)_{-n} \,\R\,u^{n-1}\,V =
  \sum_{n=0} ((L_{-n}-\Lf_{-n}) \,\R\, \HH_0\,u^{n-1}\,V
\eqno(B.27)$$
so we rewrite
$\nu (\hf^1\diamond\HH)_{-t} =  (T\diamond\HH)_{-t-1}
  - (\Tf\diamond\HH)_{-t-1} - \nu (\hf^2\diamond\HH)_{-t}$
and use
$$
  \sum_{t=1} (\hf^2\diamond\HH)_{-t} \, \R\, u^t\, V =
  \sum_{t=1}\left( (\HH\,\hf^2)_{-t} - \hf^2_{-t}\,\HH_0
  - {C_{11}\over\nu}(\hf^2 + \p \hf^2)_{-t}\right) \,\R\,u^t\,V \,.
\eqno(B.28)$$

Next note that
$$\eqalign{
  0 &= \R\,\hf^3_1\,V = \sum_{n=0} \hf^3_{-n+1} \,\R \, u^n \, V\,,
\cr\cr
  0 &= \R\,\sum_{n=1} C_{2j} (\hh^j_{-n}\,\hf^1_n
                             + \hf^1_{-n}\,\hh^j_n) \, V \cr\cr
    &= \sum_{t=0} \left( \sum_{n=1} C_{2j} (\hh^j_{-n}\,\hf^1_{-t+n}
                             + \hf^1_{-n}\,\hh^j_{-t+n}) +
    {C_{12}\over\nu} (\p \hf^1 + \hf^1)_{-t} \right) \,
    \R \, (-u)^t \, V \cr}
\eqno(B.29)$$
because by the property $(B.1)$ any positive mode generator
annihilates
the vector $V$.

Putting all this together we get for $(B.26)$:
$$\eqalign{
& \sum_{n=0}^\infty \R_{-n}\, \Big(\hh^3_0 + 2 - \kk-n\Big)
   \,\f10\, u^n \, V =
\cr\cr
&=   \sum_{t=0} \bigg(
  \hf^3_{-t+2} + (h^3_0 + 1)\hf^1_{-t+1} - \big(1-{1\over\nu}\big)
  (\p \hf^1)_{-t+1}
- {C_{11}\over\nu}(\hf^1+\hf^2+\p \hf^1 +\p \hf^2)_{-t+1}
\cr\cr
&  + (C_{2j}(\hh^j\,\hf^1)_{-t+1}
   -   C_{1j}(\hh^j\,\hf^2))_{-t+1} - C_{2j}\,\hf^1_{-t+1}\,
\hh^j_0
  + C_{1j}\, \hf^2_{-t+1}\, \hh^j_0 \Big) \R (-u)^t V \,.\cr}
\eqno(B.30)$$
With this we have completed step 2) and thus the derivation of
(4.5).
 \medskip

%%%%%%%%%%%
%
%     appendix C
%
%%%%%%%%%%%

\bigskip\noindent{\bf Appendix C. Derivation of the cohomology
equivalence (6.15)}

\medskip

 In this appendix we will demonstrate how in a
Malikov-Feigin-Fuks monomial the factors
corresponding to $w_{020}=w_{\delta-\alpha^1}$
and $w_{010}=w_{\delta-\alpha^2}$ are
cohomologically equivalent to powers of the
corresponding root vectors, i.e.,
$\he^1_{-1}$ and $\he^2_{-1}$, respectively, and hence,
accounting for the constraints, these factors can be
set equal to a constant modulo $Q$-exact terms.

For the generators of ${\tt n}_-$ we use the notation $\fb_i$,
$i$ runs
over the simple roots of the affine algebra. In particular
$\fb_0=e^3_{-1}$ and $\fb_2=f^2_0$. From the commutation
relations we have
$$
  \fb_0^{\,1+A} \, \fb_2 \,\fb_0^{\,-A}
  = \fb_0 \, \fb_2 + A\, e^1_{-1} \,.
\eqno(C.1)$$
Using the quantum constraint $\fb_0 = \{ Q, b^3_0 \}$ we get
$$
   \fb_0 \, \fb_2 = b^3_0 \,[ (c^2 \, h^2)_{-1}+ (c^3\,
e^1)_{-1}] +
   \{Q, b^3_0\,\fb_2\} \,.
\eqno(C.2)$$

As a first example consider (C.1) applied to a singular vector
$V_\mu$ such that $\la \za^2 \,,\zm \ra =A$, i.e.,
$$
  h^2_0\, V_{\mu}  =  A\, V_{\mu} \,.
\eqno(C.3)$$
Then
$$
  \fb_0^{\,1+A} \, \fb_2 \,\fb_0^{\,-A}\,\Vm \simeq
  A\, \left( e^1_{-1} + (b^3_0\,c^2_{-1})\right) \,\Vm
  =  A\, \hat e^1_{-1} \,\Vm
\eqno(C.4)$$
and substituting the constraint $ \hat e^1_{-1} = 1 +
\{Q,b^1_0\}$
we get
$$
  \fb_0^{\,1+A} \, \fb_2 \,\fb_0^{\,-A}\Vm \simeq A \, \Vm
\eqno(C.5)$$
where $\simeq$ denotes equality modulo $Q$ exact terms.

As a second example consider $\fb_0^{\,2+A} \, \fb_2^{\,2}
\,\fb_0^{\,-A} $
applied on a singular vector $\Vm$ such that
$$
  \la \za^2 \,,\zm \ra = A + 1\,.
\eqno(C.6)$$
Using twice (C.1) we can write
$$
  \fb_0^{\,2+A} \, \fb_2^{\,2} \,\fb_0^{\,-A} =
  \fb_0\,\fb_2 \left(  \fb_0\,\fb_2 + (2A+1) e^1_{-1} \right)
  + A(A+1) \, (e^1_{-1})^2 \,.
\eqno(C.7)$$
For the first term on the r.h.s. of the above using (C.2),
the constraint for $\fb_0$, and (C.6) we get
$$\eqalign{
  \fb_0\,\fb_2 \left(  \fb_0\,\fb_2 + (2A+1) e^1_{-1} \right) \,
\Vm &=
  (b^3_0 \, (c^2 \, h^2)_{-1} + \{Q, b^3_0\,\fb_2\}) \,
  \left(  \fb_0\,\fb_2 + (2A+1) e^1_{-1} \right) \, \Vm \cr\cr
  &\simeq A \, (b^3_0\,c^2_{-1}) \, (- [\fb_2,\fb_0] + (2A+1)
e^1_{-1} )
  \, \Vm \cr}
\eqno(C.8)$$
and thus
$$
  \fb_0^{\,2+A} \, \fb_2^{\,2} \,\fb_0^{\,-A} \,\Vm \simeq
   A(A+1) \,
   \left( 2\,(b^3_0\,c^2_{-1}) \, e^1_{-1}+(e^1_{-1})^2\right) \,
\Vm
   = A(A+1) \, (\hat e^1_{-1})^2 \, \Vm\,.
\eqno(C.9)$$
and substituting the constraint for $\hat e^1_{-1}$ we finally
have
$$
  \fb_0^{\,2+A} \, \fb_2^{\,2} \,\fb_0^{\,-A} \,\Vm \simeq A(A+1)
\,\Vm
\eqno(C.10)$$

More generally if $\Vm$ is such that
$$
  \la \za^2 \,,\zm \ra = (A+p-1)
\eqno(C.11)$$
then
$$
  \fb_0^{\,p+A} \, \fb_2^{\,p} \,\fb_0^{\,-A} \,\Vm
  \simeq {\Gamma(A+p)\over\Gamma(A)}  \,\Vm
\eqno(C.12)$$

We will extend  the last formula for arbitrary $p$, not
necessarily integer.
Now consider  $w_{020}$ applied to
$w_{1(0201)^t}\cdot\lambda$.
Thus we are
in the situation of (C.12) with
 $$\mu=w_{1(0201)^t}\cdot\lambda\,, \,
 A=M^2-{t+1\over\nu} \,, \,  p=M^1+{2t+1\over\nu}\,,$$
   and
 $$h^2(w_{1(0201)^t}\cdot\lambda)=M^1+M^2+{t\over\nu}-1\,, $$
thus
the condition (C.11) is satisfied and we get
$$
  V_{w_{020}w_{1(0201)^t}\cdot\lambda} \simeq
  {\Gamma(M^1+M^2+{t\over\nu})\over\Gamma(M^2-{t+1\over\nu})}\,
    V_{w_{1(0201)^t}\cdot\lambda}
\eqno(C.13)$$
Repeating this step consecutively for $t=n-1, n-2, .... ,0$ one
effectively ``wipes away'' all groups of Kac-Moody generators in
(A.1)
 corresponding to the
reflections  $w_{020}$. The powers of $\fb_1$ are replaced by
the operators (6.7) as in (6.8)  and altogether this reproduces
the
explicit expression (6.18) of the singular vector.
Collecting the numerical factors from (C.13) for
a singular vector we get, e.g., for the coefficient $N_1$
in (6.18)
$$
  N_1 = \prod_{t=1}^{m_1'-1}
  {\Gamma(\langle\lambda+\rho,\alpha^3+(t-1)\delta\rangle )
  \over\Gamma(\langle\lambda+\rho,\alpha^2-t\delta\rangle )}\,.
\eqno(C.14)$$
The same formula with $\za^2\,,m_1'$ replaced by  $\za^1\,,m_2'$
reproduces up
to a sign the constant $N_2$ in (6.19). The constant $N'_1$
(which appear in the case $M^1=-m_1+{m_1'-1\over \nu}$) is
provided by the
same expression (C.14) with $\za^2$ and $\za^3$ interchanged,
while $N'_2$
(for $M^2=-m_2+{m_2'-1\over \nu}$) is
obtained from the corresponding expression of $N_2$ with $\za^1$
and $\za^3$
interchanged.

To obtain (6.20) one has to use
that the reflection in (2.12c) can be also rewrittten as
$w_{\beta}=
w_{12} w_{(1012)^{n}}
w_{1} = w_{12} w_{(0102)^{n}} w_{1}\,.$ Then one has
to transform the corresponding MFF monomial as above. The
resulting overall
constant $N_3$ in (6.20) reads ( again up to a sign )
$$
  N_3 = \prod_{t=1}^{m_3'-2}
  {\Gamma(\langle\lambda+\rho,\alpha^2+(t-1)\delta\rangle )
  \over\Gamma(\langle\lambda+\rho,-\alpha^1-t\delta\rangle )}\,,
\eqno(C.15)$$
while $N'_3$ (for $M^3=-m_3+{m_3'-2\over \nu}$) is obtained from
(C.15) interchanging $\za^2$ and $-\za^1$ .

%%%%%%%%%%%
%
%     appendix D
%
%%%%%%%%%%%

\bigskip\noindent{\bf Appendix D.  Analytic continuation
of ${\cal O}_\lambda$ and an explicit example of
a singular vector.  }
\medskip

In this appendix we will illustrate on a concrete example how
one can work with the formal expressions (6.18-20)
and get explicit results.

Consider the expansion (6.17) of ${\cal O}^{(i)}_\lambda$
in  decreasing powers of $W_{-1}$,
keeping terms up to $k=2$
 (for notational clarity we will
do the case $i=1$;
for short in this appendix we absorb the normalization
$\aw$ into the current $W(z)$.)
$$
  {\cal O}^{(1)} = K\,\left(
  (W_{-1})^p + A\,(W_{-1})^{p-1} L_{-1} + (W_{-1})^{p-2}
  \left( B\,L_{-1}^2 + C\,L_{-2} + D\,W_{-2} \right) + \dots
\right)
\eqno(D.1)$$
The dots stand for terms of lower order in $W_{-1}$.
For short denote the parameters $(M^1,M^2)$ of the weight
$\lambda$ by $(p,q)$.
The overall normalization is
$$
  K(p,q) =
  (-1)^p {\Gamma(q-{1\over\nu}) \over \Gamma(p+q-{1\over\nu})}
\eqno(D.2)$$
while the coefficients
$A$, $B$, etc., are polynomials in the parameters:
$$\eqalign{
  A(p,q)&= {p(p+2q-{3\over\nu})\over 6\nu}
\cr\cr
  B(p,q)&= {p(p-1) \left( 4q^2 + (4p-{12\over\nu})q
           + p^2 - ({6\over\nu}+3)p + 9k^2 - 3 \right) \over
72\nu^2}
\cr\cr
  C(p,q)&= {p(p^2-1) \left(-10q^2+({30\over\nu}-10p)q
   - 4 p^2 + {15\over\nu}p - {20\over\nu^2} +6 \right) \over
180\nu}
\cr\cr
  D(p,q)&= {p(p-1)(p+{1\over\nu}+1)({3\over\nu}-p-2q)\over 12}
\cr\cr
  etc. &\quad
\cr}
\eqno(D.3)$$
The derivation of (D.3) is straightforward though lengthy.
One starts from the definition (6.4)
assuming that $M^1=p$ is an integer
$$
  (-1)^{M^1} \, {\cal O}^{(1)}_\lambda =
 {\cal L}^{(1)}_{p,p-1}\dots{\cal L}^{(1)}_{2,1}{\cal
L}^{(1)}_{1,0}
 +
   \sum_{q=2}^p
   \underbrace{ {\cal L}^{(1)}_{p,p-1}\dots{\cal L}^{(1)}_{q+1,q}
}
             {\cal L}^{(1)}_{q,q-2}
   \underbrace{ {\cal L}^{(1)}_{q-2,q-3}\dots{\cal L}^{(1)}_{1,0}
}
 + \dots
\eqno(D.4)$$
To obtain the expansion up to the order displayed in (D.1) we
need only the two terms in (D.4) and the commutator
$[L_{-1},W_{-1}]=-W_{-2}$.
We also make multiple use of the formula
$$
  \sum_{q=0}^p {q\choose k} = {p+1\choose k+1}    \,.
\eqno(D.5)$$
Since the coefficients (D.3) are polynomials in $p$ we can
use the expansion of ${\cal O}^{(1)}$ in powers of $W_{-1}$
to define it for arbitrary complex $M^1=p$.
Of course, in analogy with [18] we have to enlarge
the algebra to contain arbitrary powers of $W_{-1}$.

Next we have to consider compositions $\tilde{\cal O}{\cal O}$,
where $\tilde{\cal O}$ has the expansion (D.1) as ${\cal O}$
but with coefficients $\tilde A=A(\tilde p,\tilde q)$,
$\tilde B=B(\tilde p,\tilde q)$, etc. Up to the same order
we get (again we need only $[L_{-1},W_{-1}]=-W_{-2}$)
$$\eqalign{
&  \tilde{\cal O}\, {\cal O} =
  \tilde K\,K\, (W_{-1})^{\tilde p + p -2} \cdot
\cr\cr
  &\cdot\left( W_{-1}^2 + (A+\tilde A)\, W_{-1}\,L_{-1} +
   (B+\tilde B + A\tilde A)\,L_{-1}^2 + (C+\tilde C)L_{-2}
   + (D+\tilde D - p\tilde A)W_{-2} + \dots\right)
\cr}
\eqno(D.6)$$

Now we turn to
the simplest example with $\lambda$ such that
 $M^1=1-{1\over\nu}$, ($M^2=q$ arbitrary).
The singular vector corresponding to $\beta=\delta+\alpha^1$ is
$$
  {\cal O}^{(1)}_{w_{0201}\cdot\lambda}\,
  {\cal O}^{(1)}_\lambda \, V_\lambda  \,.
\eqno(D.7)$$
In (D.6) we have to set $p=1-{1\over\nu}$, $\tilde
p=1+{1\over\nu}$
and $\tilde q=q-{1\over\nu}$ obtaining
$$\eqalign{
  A+\tilde A &= {1\over 3\nu} (1 - {4\over\nu} + 2q )
\cr\cr
  B + \tilde B + A\tilde A &=
  {4q^2 + (4 - {16\over\nu})q + {7\over\nu^2} - {8\over\nu} +1
  \over 36 \,\nu^2}
\cr\cr
  C + \tilde C &=
  { -q^2 + ({4\over\nu}-1) q - {4\over\nu^2} + {2\over\nu}
  \over 3 \,\nu^3}
\cr\cr
  D + \tilde D - p \tilde A &=
  {({3\over\nu}+1)({4\over\nu}-1-2q)\over 6\,\nu}
\cr}
\eqno(D.8)$$
Moreover the full analysis shows that the other coefficients
vanish, i.e., the dots in the r.h.s. of (D.6) for these
particular parameters can be removed, and we get a finite
polynomial expression with integer powers of the ${\cal W}$
algebra generators for the ${\cal P}_{\beta,\lambda}$.

In this particular example one can obtain the singular vector in
a
much simpler way and the above exercise becomes merely a check
that
our algorithm is correct. The class
of weights $\lambda$
such that $M^1=m-{1\over\nu}$ (or $M^2=m-{1\over\nu}$),
for some $m=0,1,2,\dots$, maps into the class
of weights integral along the first (or second) root under the
following generalization of the duality tranformation of [14]
$$
  \nu\mapsto\overline\nu={1\over\nu}\,,\quad
  M^1\mapsto\overline M^1=1-\nu(M^2-1)\,,\quad
  M^2\mapsto\overline M^2=1-\nu(M^1-1)\,.
\eqno(D.9)$$
In our particular example (where $m=1$) we have
$$
  \overline M^1=1+\nu-\nu M^2\,,\quad
  \overline M^2=2
\eqno(D.10)  $$
and the singular vector (D.7) is equal up to an overall constant
to ${\cal O}^{(2)}_{\overline\lambda}\, V_{\overline\lambda}$.
Recall that for $\overline\lambda$ with (D.10) the operator
${\cal O}^{(2)}_{\overline\lambda}$ is an ordinary polynomial
in the generators $W_{-n}$ and $L_{-n}$ and no analytic
continuation is necessary, namely
$$
  {\cal O}^{(2)}_{\overline\lambda}
  = \overline{\cal L}^{(2)}_{2,1}\,\overline{\cal L}^{(2)}_{1,0}
    + \overline{\cal L}^{(2)}_{2,0} \,
\eqno(D.11)$$
where for short $\overline{\cal L}$ stands for ${\cal L}$ with
$\overline\lambda$.

\bigskip
\bigskip\noindent{\bf Acknowledgements}
\medskip

 We are indebted to L. Bonora and especially to V.K. Dobrev for
useful discussions and to A. Koubek and C.J. Zhu for patient help
with MATHEMATICA.  V.B.P.  acknowledges the hospitality  of the
Arnold Sommerfeld Institute for Mathematical Physics, TU
Clausthal.   A.Ch.G and V.B.P.  acknowledge the financial support
and  the hospitality  of INFN, Sezione di Trieste and SISSA,
Trieste. A.Ch.G. was supported also by the Humboldt foundation.

This work was supported in  part by the Bulgarian Foundation for
Fundamental Research under contract $\phi -11 - 91. $
\bigskip
\bigskip

\bigskip
{\bf Note Added}
\smallskip

\medskip
After this work was essentially written the paper [36] appeared
in hep-th.  The main result of this paper is to give a rigorous
meaning of the analytic continuation of the expressions ${\cal
O}^{(i)}_\lambda$ to complex
$\langle\lambda+\rho,\alpha^i\rangle$ and hence of the general
${\cal W}_3$ singular vectors. In this respect it is
complementary to our main emphasis.  Nevertheless we have left
the explicit example in appendix D which demonstrates that the
method of Kent is algorithmic and can be used for concrete
calculations just as the fusion method of [24].

In [36] there is also a proof of the important fact that there is
exactly one ${\cal W}_3$ algebra singular vector for every weight
-- something we have only conjectured.

Finally we should point out that though the author of [36] argues
that the Verma modules with pseudo-singular vectors occuring are
exhausted by the weights  ${1\over\nu} = m = M^1 = M^2\,, m\in
\dN\,,$, in fact there are other possibilities, demonstrated for
the case $\nu=1$ already in [32].

%%%%%%%%%%
%
%  references
%
%%%%%%%%%%

\bigskip\noindent{\bf REFERENCES}
\medskip

\item{[1]} %[DS]
 V. Drinfeld and V. Sokolov, {\it J. Sov. Math.} {\bf 30} (1984)
1975.

\item{[2]}% [Bel]
  A.A. Belavin, in: N.Kawamoto, T. Kugo (Eds.), ``Quantum string
  theory'', Springer Proc. in Phys., Vol. 31, p. 132,
  Springer Verlag, Berlin, 1988.

\item{[3]} %[BFFOW]
   J. Balog, L. Feh\'er, P.Forgac, L. O$'$ Raifeartaigh and A. Wipf,
 {\it Phys. Lett.} {\bf B244} (1990) 435,
 {\it Ann. Phys.} {\bf B203} (1990) 76.

\item{[4]} %[BTD]
 F.A. Bais, T. Tjin and P. van Driel
    {\it Nucl. Phys.} {\bf B357} (1991) 632.

\item{[5]}% [FORTW]
  L. Feh\'er,  L. O$'$ Raifeartaigh,
O. Ruelle, I.Tsutsui and A. Wipf,
 {\it Phys. Rep.}  {\bf 222} (1992) 1.

 \item{[6]}% [ F Luk]
V.A.  Fateev and S.L. Lukyanov, {\it Int. J. Mod. Phys.} {\bf A3}
(1988) 507.

\item{[7]} %[BO]
   M. Bershadsky and H. Ooguri, {\it Comm. Math. Phys.} {\bf
  126} (1989) 49.

\item{[8]}% [FF]
   B. Feigin and E. Frenkel, {\it Phys. Lett.} {\bf B246} (1990) 75.

\item{[9]} %[B]
   M. Bershadsky,  {\it Comm. Math. Phys.} {\bf 139 } (1991) 71.

\item{[10]} %[BT1]
 J. de Boer and T. Tjin,  {\it Comm. Math. Phys.} {\bf
158 } (1993) 485.

\item{[11]}% [BT2]
 J. de Boer and T. Tjin,  {\it Comm. Math. Phys.} {\bf
160 } (1994) 317.

\item{[12]} %[W]
  G.M.T. Watts,
 {\it Nucl. Phys.} {\bf B326} (1989) 648.

\item{[13]} %[FKW]
   E. Frenkel, V. Kac and M. Wakimoto, {\it Comm. Math. Phys.} {\bf
147 } (1992) 295.

\item{[14]}% [FGPP]
   P. Furlan, R. Paunov, A.Ch. Ganchev and V.B. Petkova,
   {\it Phys. Lett.} {\bf B267} (1991) 63;
    {\it Nucl. Phys.} {\bf B394} (1993) 665.

\item{[15]} %[KK]
 V.G. Kac and D.A. Kazhdan, {\it Adv. Math.} {\bf 34} (1979) 97.

\item{[16]}% [MFF]
   F.G. Malikov, B.L. Feigin and D.B. Fuks, {\it Funct. Anal.
   Prilozhen.} {\bf 20}, no.  2 (1987) 25.

\item{[17]}% [GP2]
    A.Ch. Ganchev and V.B. Petkova,
   {\it Phys. Lett.} {\bf B318} (1993) 74.

\item{[18]} %[K]
    A. Kent, {\it Phys. Lett.} {\bf B273} (1991) 56.

\item{[19]}% [GP]
   A.Ch. Ganchev and V.B. Petkova,
{\it Phys. Lett.} {\bf B293}   (1992) 56.

\item{[20]}% [FGP]
    P. Furlan, A.Ch. Ganchev and V.B. Petkova,
    {\it Phys. Lett.} {\bf B318} (1993) 85.

\item{[21]}% [Z]
A.B. Zamolodchikov, {\it Theor. Math. Phys.} {\bf 65} (1988) 1205.

\item{[22]}% [P]
   A. Polyakov, {\it Int. J. Mod. Phys.}  {\bf A5} (1990) 833.

\item{[23]}% [BW]
  P. Bowcock and G.M.T. Watts,
 {\it Phys. Lett.} {\bf B297} (1992) 282.

\item{[24]}% [BFIZ]
     M. Bauer, Ph. Di Francesco, C. Itzykson and J.-B. Zuber,
    {\it Phys. Lett.} {\bf B260} (1991) 323;
    {\it Nucl. Phys.} {\bf B362} (1991) 515.

\item{[25]}% [FM]
  B.L. Feigin and  F.G. Malikov, Integral intertwining operators and
complex powers of differential ($q$ - difference) operators, preprint
RIMS -894 (1992), {\tt hep-th 9306137},
to appear in Advances in Sov. Math.

\item{[26]} %[D]
V.K. Dobrev, Multiplet classification of the indecomposable
highest weight modules over \ affine Lie algebras and invariant
differential operators: the $A_l^{(1)}$ example, ICTP preprint
IC/85/9.

\item{[27]}%[BBSS]
       F.A. Bais, P. Bouwknegt, M. Surridge and K. Schoutens,
    {\it Nucl. Phys.} {\bf B304} (1988) 348.

\item{[28]}%[T]
K. Thielemans,
{\it Int. J. of Mod. Phys.} {\bf C2} (1991) 787.

\item{[29]}%[FZ]
 V.A. Fateev and A.B. Zamolodchikov,
 {\it Nucl. Phys.} {\bf B280} (1987) 644.

\item{[30]} %[M]
 S. Mizoguchi, {\it Phys. Lett.} {\bf B222} (1989) 226.

\item{[31]}% [BMP1]
  P. Bouwknegt, J. McCarthy and K. Pilch,
  {\it Lett. Math. Phys.} {\bf 29} (1993) 91.

\item{[32]} %[BMP2]
P. Bouwknegt, J. McCarthy and K. Pilch, On the BRST structure of $\W_3$
gravity coupled to $c=2$ matter, preprint USC-93/14, ADP-93-203/M17.

\item{[33]}% [dVD]
  K. de Vos and P. Van Driel, The Kazhdan-Lusztig conjecture for
finite W-algebras, DAMTP-93-66, UCLA/93/TEP/46, hep-th/9312016

\item{[34]}% [BG]
  A. Bilal and J.-L. Gervais,
 {\it Nucl. Phys.} {\bf B318} (1989) 579.

\item{[35]}% [BPT]
  Z. Bajnok, L. Palla and G. Tak\'acs,
 {\it Nucl. Phys.} {\bf B385} (1992) 329.

\item{[36]}
  Z. Bajnok, Singular vectors of the $W A_2$ algebra, Cambridge
preprint,
  hep-th/9403032.

\bye